\documentclass[aps,prb, amsmath,amssymb,twocolumn,superscriptaddress]{revtex4-2}

\usepackage{graphicx}
\usepackage{bm}
\usepackage{hyperref}
\usepackage{dsfont}
\usepackage{color,soul}
\renewcommand{\i}{\bm{i}}
\renewcommand{\v}{\bm{v}}
\renewcommand{\j}{\bm{j}}
\def\Tr{\mathop{\mathrm{Tr}}}
\newcommand{\q}{\boldsymbol{q}}
\newcommand{\x}{X}
\usepackage{ulem}
\usepackage{xcolor}

\usepackage{xspace}
\newcommand{\eg}[0]{e.g.\@\xspace}
\newcommand{\ie}[0]{i.e.\@\xspace}

\usepackage{ifthen}
\newboolean{draft}

\setboolean{draft}{true}   

\ifthenelse{\boolean{draft}}{%
}{%
}%
\ifthenelse{\boolean{draft}}{%
}{%
}%
\ifthenelse{\boolean{draft}}{%
}{%
}%
\ifthenelse{\boolean{draft}}{%
}{%
}%
\ifthenelse{\boolean{draft}}{%
}{%
}%
\ifthenelse{\boolean{draft}}{%
}{%
}%
\ifthenelse{\boolean{draft}}{%
  \newcommand{\rem}[1]{{\textcolor{red}{\sout{#1}}}}%
}{%
  \newcommand{\rem}[1]{}%
}%

\begin{document}

\title{Phases and Exotic Phase Transitions\\ of a Two-Dimensional Su-Schrieffer-Heeger Model}
\author{Anika  G\"otz}
\affiliation{\mbox{Institut f\"ur Theoretische Physik und Astrophysik,
    Universit\"at W\"urzburg, 97074 W\"urzburg, Germany}}
\author{Martin  Hohenadler}
\affiliation{\mbox{Institut f\"ur Theoretische Physik und Astrophysik,
    Universit\"at W\"urzburg, 97074 W\"urzburg, Germany}}
\author{Fakher F. Assaad}
\affiliation{\mbox{Institut f\"ur Theoretische Physik und Astrophysik,
    Universit\"at W\"urzburg, 97074 W\"urzburg, Germany}}
\affiliation{W\"urzburg-Dresden Cluster of Excellence ct.qmat, Am Hubland, 97074 W\"urzburg, Germany}

\date{\today}

\begin{abstract}
  We study a Su-Schrieffer-Heeger electron-phonon model on a square lattice by
  means of auxiliary-field quantum Monte Carlo simulations. The addition of a
  symmetry-allowed interaction permits analytical integration over the phonons
  at the expense of discrete Hubbard-Stratonovich fields with imaginary-time correlations.
  Using single-spin-flip  and  global updates, we investigate the phase diagram at
  the O(4)-symmetric point as a function of hopping $t$ and phonon
  frequency $\omega_0$. For $t=0$, where electron hopping is boson assisted, the model maps
  onto an unconstrained $\mathbb{Z}_2$ gauge theory. A key quantity is the
  emergent effective flux per plaquette, which equals $\pi$ in the
  assisted-hopping regime and vanishes for large $t$.  
  Phases   in the former regime can be 
  understood in terms of instabilities of emergent Dirac fermions. Our results
  support a direct and continuous transition between a
  $(\pi,0)$ valence bond solid (VBS) and an antiferromagnetic (AFM) phase with
  increasing $\omega_0$. For large $t$ and small $\omega_0$, 
  we find finite-temperature signatures, a disordered pseudogap  phase, 
   of a previously reported $(\pi,\pi)$ VBS
  ground state related to a nesting instability.  
   With increasing $\omega_0$, AFM order again emerges.
\end{abstract}

\maketitle

\section{Introduction}\label{sec:introduction}

One of the most fundamental interaction channels in the solid state is the
coupling between lattice vibrations (phonons) and conduction electrons.  In a
Fermi liquid with a coherence temperature orders of magnitude
greater than the Debye frequency, electron-phonon coupling leads to a retarded
and net attractive interaction.  The Cooper instability of Fermi
surfaces promotes superconductivity \cite{Cooper56,Bardeen57}. However,
electron-phonon coupling does not always lead to superconductivity. Notably,
in one dimension (1D), where $2k_\text{F}$ nesting is
generic, it triggers a Peierls charge-density-wave (CDW) instability. In two-dimensional (2D) systems,
nested Fermi surfaces lead to, \eg, $(\pi,\pi)$ valence bond solid (VBS) or
antiferromagnetic (AFM) order \cite{Cai21,Goetz22}.

Here, we investigate if a symmetry-allowed generalization of the
Su-Schrieffer-Heeger (SSH) model \cite{Su79} on a square
lattice can host exotic phases and quantum phase transitions. This question has
previously been answered affirmatively for a spinless 1D SSH model, which exhibits
instances of 1D deconfined quantum criticality \cite{Senthil04b,Jiang19,Mudry19,Weber20}.

\begin{figure}[h]
  \includegraphics[width=1\linewidth]{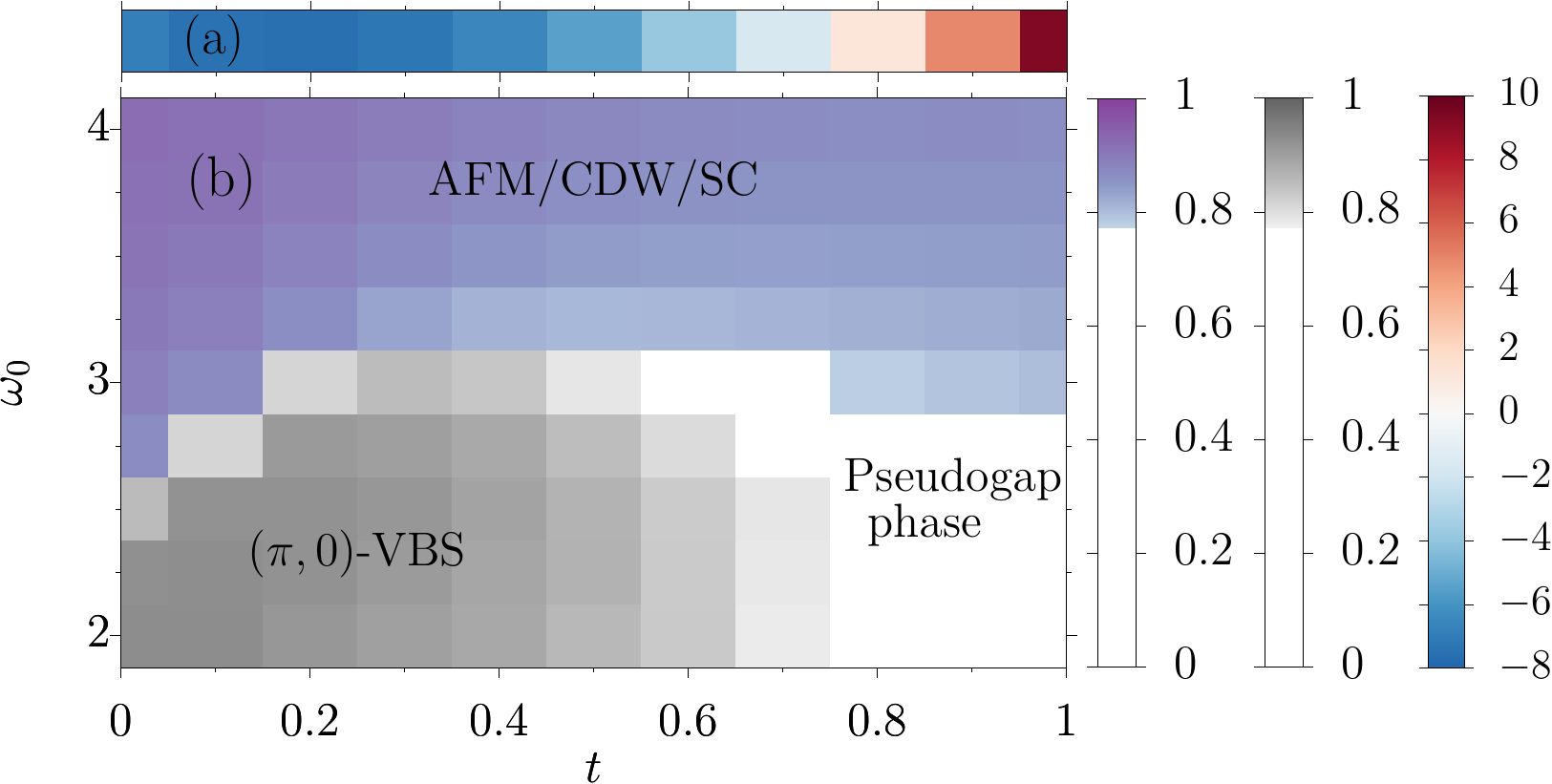}
  \caption{\label{fig:phases}Sketch of the phase diagram of
    Hamiltonian~(\ref{eq:SSH}) as a function of electronic hopping $t$ and
    phonon frequency $\omega_0$, based on QMC simulations. It features
    antiferromagnetic (AFM), valence-bond solid (VBS), and pseudogap phases.
    Due to the $\mathrm{O}(4)$ symmetry of the model the AFM phase is degenerate
    with charge-density-wave (CDW) ordering and $s$-wave superconductivity (SC).
    At  small  $t$, we  observe the spontaneous generation of a $\pi$ flux in
    each plaquette. (a) Average flux $\langle \hat{\Phi}_{\bm{i}}\rangle$ as a function of $t$ at $\omega_0=2.0$, (b)
    correlation ratio of the spin susceptibility $R_{\chi,S}$ at $(\pi,\pi)$ (lilac scale), and the VBS susceptibility $R_{\chi,D(\pi,0)}$ at
    $(\pi,0)$ (gray scale). Results in (a) and (b) are for $L=\beta=8$. 
    Here, $\tilde{g}=\sqrt{2/k}g=2$ and $\lambda=0.5$.
    Since the correlation ratios in (b) are limited to one lattice size and the cutoff of the color scale is 
    arbitrary, the numerical values of the phase boundaries are only a rough estimation.
   }
\end{figure}
  
From the original perspective of electron-phonon coupling, the phonon-mediated
modulation of the direct hopping $t$ in the SSH model has to be small
\cite{Su79}.  However, we can take a more general view by also considering the
regime of small (vanishing) $t$, where electronic hopping is partially
(exclusively) phonon assisted. Related fermion-boson models have been put forward to describe
the motion of holes in an antiferromagnetic background \cite{Edwards06}\footnote{Note,
 however, that the so-called Edwards model differs in the sense
  that hopping from $\i$ to $\j$ is possible only by
  emitting a bosonic mode that can be re-absorbed when hopping from $\j$ to
  $\i$.}.
Similar to Ref.~\cite{Seki18}, our model includes an additional electronic
interaction term  corresponding to  the square of the hopping. 
Both models,  with and without symmetry-allowed interaction term,  do not  suffer  from the 
negative sign problem in auxiliary-field quantum Monte Carlo (QMC) simulations. 
However, the additional term allows  us  to integrate out the fermions at the expense of a  
retarded interaction between  discrete  auxiliary fields. This results in a reduced  autocorrelation 
time in the studied parameter regimes, as compared to the direct sampling of the phonon fields. 
We used an implementation of the finite-temperature auxiliary-field QMC algorithm
\cite{Blankenbecler81,Hirsch82,White89,Evertz08} from the Algorithms for Lattice
Fermions (ALF) package \cite{ALF22}. 

The main results are summarized by the very rich phase diagram in
Fig.~\ref{fig:phases}.   At  $t=0$, our  model maps  onto an unconstrained 
$\mathbb{Z}_2$ lattice gauge theory \cite{Assaad16,Prosko17,Gazit18}.   
In this limit, a $\pi$ flux  per  plaquette  and  associated  Dirac  fermions 
emerge. Importantly, the $t=0$ physics is adiabatically connected to a $t>0$
region where the flux  remains  negative, in which we observe dynamically
generated mass terms corresponding to $(\pi,0)$ VBS and AFM phases. The latter
are separated by an apparently direct and continuous phase transition
interpreted as a deconfined quantum critical point (DQCP) \cite{Senthil04b,Senthil06}. 
At large $t$, the $\pi$ flux vanishes and previous studies
\cite{Xing21,Cai21,Goetz22,cai2022robustness,Feng22} suggest the ground
state at small $\omega_0$ to be a $(\pi,\pi)$ VBS state. At the temperatures
considered here, we instead observe a pseudogap phase of fluctuating dimers.

The paper is organized as follows. In Sec.~\ref{sec:model-symmetries}, 
we define the model and discuss its symmetries and limiting cases, as well as
previous work. In Sec.~\ref{sec:method}, we describe our numerical method. In
Sec.~\ref{sec:results}, we present our numerical results, followed by a
discussion and conclusions in Sec.~\ref{sec:conclusion}. We provide four appendixes with further
details about the method and additional data, respectively.

\section{Model and Symmetries}\label{sec:model-symmetries}

\subsection{Model}

A generic SSH-type Hamiltonian \cite{Su79,Su80} with
Einstein phonons takes the form 
\begin{equation}\label{eq:SSH_generic}
  \hat{H}=   \sum_b   \left( -t  + g  \hat{X}_b \right)  \hat{K}_b  + \sum_b \left( \frac{1}{2m} \hat{P}_b^2  +  \frac{k}{2}   \hat{X}_b^2 \right) \,.
 \end{equation}
 The first term describes fermion hopping and fermion-phonon coupling
 on bonds $b=\langle \i,\j \rangle$ connecting nearest-neighbor sites $\i$ and $\j$, with the hopping operator
\begin{equation}\label{eq:kin}
  \hat{K}_b=\sum_\sigma \hat{c}_{\i,\sigma}^{\dagger}\hat{c}_{\j,\sigma}^{\phantom{\dagger}}+\mathrm{H.c.}=\sum_{\i,\j, \sigma } \hat{c}^{\dagger}_{\i,\sigma }
  \left(K_b\right)_{\i,\j} \hat{c}^{\phantom{\dagger}}_{\j,\sigma } \,.
\end{equation}
Here, $\left(K_b\right)_{\i,\j}=1$ if $\bm{i}$ and $\bm{j}$ are nearest
neighbors and $0$ otherwise. The operator
$\hat{c}_{\i,\sigma}^{\dagger}$ creates a fermion in a Wannier state centered at
site $\i$ and with $z$-component of spin equal to $\sigma$. We will keep the
notation general enough to allow for the case of $N$ fermion flavors. However, all numerical results will be
for $N=2$, corresponding to electrons with spin $\frac{1}{2}$. The strength of the bare
electron hopping is set by the hopping integral $t$, whereas $g$ determines the
electron-phonon coupling, which modulates the electronic hopping.
The second term in Eq.~(\ref{eq:SSH_generic}) describes bond phonons modeled as
harmonic oscillators with position operators $\hat{X}_b$, momentum operators
$\hat{P}_b$, and frequency $\omega_0^2=k/m$.

Here, we study the slightly different Hamiltonian
\begin{equation}\label{eq:SSH}
\hat{H} =   \sum_b (- t + g \hat{X}_b ) \hat{K}_b   - \lambda\sum_b  \hat{K}_b ^2  + \sum_b \left( \frac{1}{2m} \hat{P}_b^2 +  \frac{k}{2}  \hat{X}_b^2 \right)
\end{equation}
on a square lattice with $N_s = L\times L$ sites. Compared to
Eq.~(\ref{eq:SSH_generic}), we include an additional electronic interaction
$-\lambda \hat{K}_b^2$ to complete the square and facilitate integration over
the phonons, similar to recent work on the Hubbard-Holstein model
\cite{Seki18}.  The additional term  does not  alter the symmetries of the model
and will therefore also be dynamically generated.

\subsection{Symmetries}\label{sec:symm}

For half-filling and a bipartite lattice, the Hamiltonian in Eq.~(\ref{eq:SSH}) is
invariant under the partial particle-hole transformation [here,  $\bm{M}=(\pi,\pi)$]
\begin{equation}\label{eq:partial-ph}
\hat{P}^{-1}_{\sigma} \hat{c}_{\i,\sigma'}^{\dagger} \hat{P}_\sigma =
  \delta_{\sigma,\sigma'} e^{i \bm{M} \cdot \i }
  \hat{c}_{\i,\sigma'}^{\phantom{\dagger}} + \left(1-\delta_{\sigma,\sigma'}
  \right) \hat{c}_{\i,\sigma'}^{\dagger}\,.
\end{equation}
The fermion parity on site $\i$ is given by
\begin{equation}\label{eq:par}
\hat{p}_{\i} = \prod_{\sigma=1}^{N} \left( 1- 2\hat{n}_{\i,\sigma} \right)\,,
\end{equation}
where $\hat{n}_{\i,\sigma}=\hat{c}_{\i,\sigma}^{\dagger}
\hat{c}_{\i,\sigma}^{\phantom{\dagger}}$ is the fermion number operator.
The parity changes sign under transformation~(\ref{eq:partial-ph}) and can be used to detect a spontaneous
breaking of particle-hole symmetry. Because the parity is an Ising-type
order parameter, $\hat{p}_{\i}^2 = 1 $, it supports order at finite temperatures
in the 2D case considered.

Our model further exhibits an $\mathrm{O}(2N)$ symmetry on bipartite
lattices \cite{Cai21,Goetz22}. 
To prove this, we use the Majorana representation for the fermions \cite{Assaad16,Beyl17},
\begin{equation}
\hat{c}_{\i,\sigma}^{\dagger} = \frac{1}{2} \left( \hat{\gamma}_{\i,\sigma,1} - i \hat{\gamma}_{\i,\sigma,2} \right) \,.
\end{equation}
With a canonical transformation $\hat{c}_{\i}^{\dagger} \rightarrow i
\hat{c}_{\i}^{\dagger}$ on one sublattice, the hopping operator can be written as
\begin{equation}\label{eq:Kb}
\hat{K}_b =   \sum_{\sigma =1}^{N}
	  \left( \hat{c}^{\dagger}_{\i,\sigma}
            \hat{c}^{\phantom{\dagger}}_{\j,\sigma}    + \text{H.c.} \right) = \frac{i}{2} \sum_\sigma \sum_{\alpha=1}^2  \hat{\gamma}_{\i,\sigma,\alpha} \hat{\gamma}_{\j,\sigma,\alpha}\,,
\end{equation}
thereby revealing the $\mathrm{O}(2N)$ symmetry.
For the case $N=2$ considered here,
the infinitesimal generators of the $\mathrm{O}(4)$ symmetry are the spin operators
$\hat{\bm{S}}_{\i}=(\hat{S}^x_{\i},\hat{S}^y_{\i},\hat{S}^z_{\i})$
and the Anderson pseudospin operators
$\hat{\bm{\eta}}_{\i}$ \cite{Anderson58}, given by 
\begin{equation}
\hat{S}_{\i}^{\alpha}= \frac{1}{2} \sum_{\sigma,\sigma'} \hat{c}_{\i,\sigma}^{\dagger} (\tau^{\alpha})_{\sigma,\sigma'} \hat{c}_{\i,\sigma'}, \quad 
\hat{\bm{\eta}}_{\i} = \hat{P}_{\uparrow}^{-1}\hat{\bm{S}}_{\i}\hat{P}_{\uparrow}\,.
\end{equation}
Here,  $\tau^{\alpha}$ is a Pauli matrix with $\alpha=x,y,z$. 
The components of $\hat{\bm{S}}_{\i}$ and $\hat{\bm{\eta}}_{\i}$ fulfill the Lie algebra of the $\mathrm{SU}(2)$ group,
$[\hat{S}_{\i}^{\alpha}, \hat{S}_{\j}^{\beta}] =i \delta_{\i,\j} \sum_n
\varepsilon_{\alpha\beta\gamma} \hat{S}_{\i}^{\gamma}$ ($\varepsilon_{\alpha\beta\gamma}$ is the Levi-Civita
symbol), and commute among each other, $[\hat{S}_{\i}^{\alpha},\hat{\eta}_{\j}^{\beta}]=0$. The Lie algebra of the global $\mathrm{O}(4)$ symmetry can be
interpreted as $\mathrm{O}(4)=\mathrm{SU}(2)\times \mathrm{SU}(2) \times \mathbb{Z}_2
$,  where the additional $\mathbb{Z}_2$ symmetry corresponds to the partial
particle-hole symmetry \cite{Assaad16}.  This implies that a potential AFM
phase is degenerate with a CDW and an $s$-wave
superconductor (SC). More specifically applying the partial particle-hole 
transformation maps the spin correlator at 
ordering wave vector $\bm{q}$ onto the density correlator
at the same wave vector and onto the superconducting correlator
at the shifted wave vector $\bm{q}-\bm{M}$.
If the parity operator acquires a nonzero expectation
value due to spontaneous breaking of the particle-hole symmetry, either the spin
or the charge sector is explicitly chosen. However due to the nature of QMC
simulations the expectation value of the parity $\langle \hat{p}_{\bm{i}}\rangle=0$ 
is always zero and the correlation function of the AFM/CDW/SC are exactly degenerate 
in the whole phase space. In order to measure a finite value of  the parity,  a small,
but finite $\mathrm{O}(4)$-symmetry-breaking term is necessary, such as a Hubbard-$U$ term.

\subsection{Limiting cases}\label{sec:limit}

\subsubsection{Adiabatic limit $\omega_0=0$}

\begin{figure}[t]
  \includegraphics[width=0.8\linewidth]{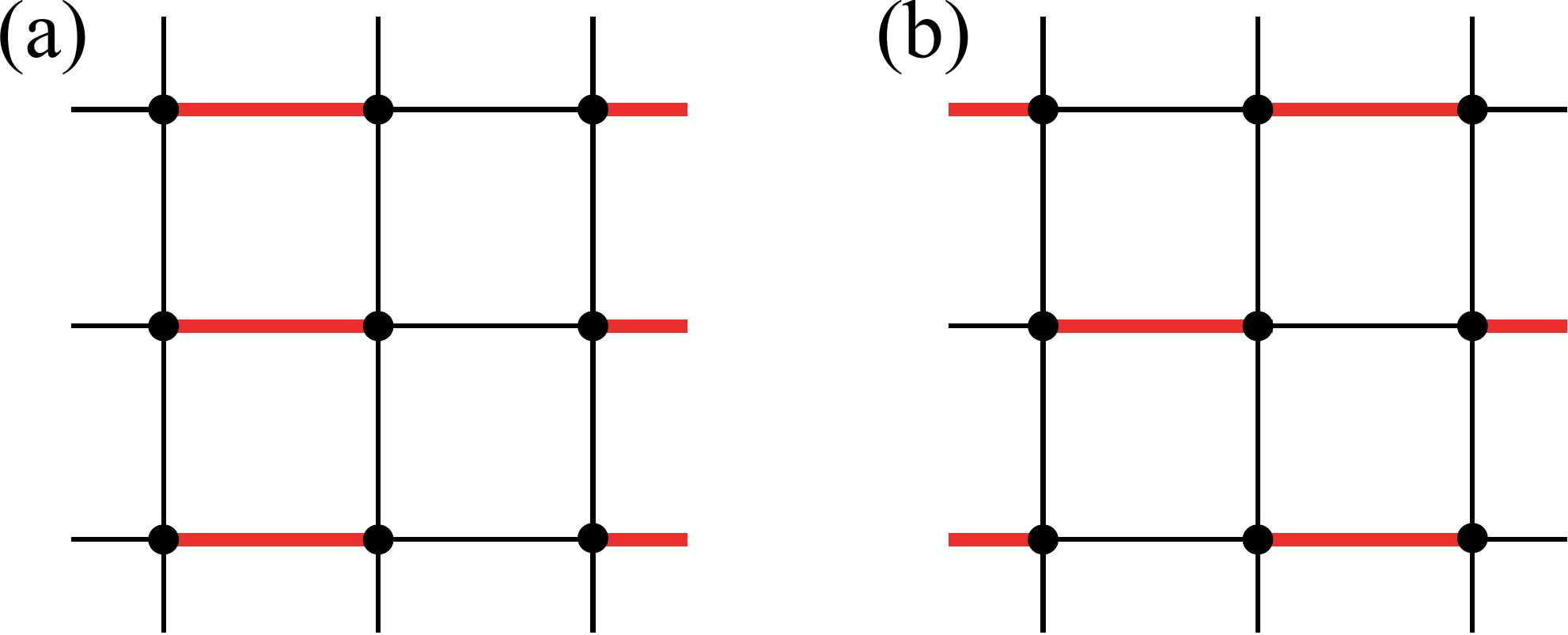}
  \caption{\label{fig:VBS}VBS phase with (a) $(\pi,0)$ ordering and (b) $(\pi,\pi)$ ordering with strong effective hopping
  $(-t+gX_b)$ on thick, red bonds and weak effective hopping on thin, black bonds. Rotating the patterns by multiples of $\frac{\pi}{2}$
  yields degenerate states [for the $(\pi,0)$ VBS a rotation by $\frac{\pi}{2}$ results in a degenerate $(0,\pi)$ VBS state].
   }
\end{figure}

In the adiabatic limit $\omega_0=0$, the phonons can be treated classically since quantum
fluctuations are frozen out. The problem reduces to finding the phonon field
configuration that minimizes the free energy of the mean-field Hamiltonian
\begin{equation}
\label{eq:SSH0}
\hat{H} = \sum_b (-t + g \x_b) \hat{K}_b + \frac{k}{2} \sum_b \x_b^2 - \lambda \sum_b \hat{K}_b^2 \, .
\end{equation}
The fields $\x_b$ are defined as the eigenvalues of the position operator, $\hat{X}_b|x\rangle=\x_b|x\rangle$.
In Fig.~\ref{fig:VBS} two possible patterns for the phonon fields are shown.
For $\lambda=0$, this model has been extensively studied  (see Sec.~\ref{sec:previous-work}).
 
\subsubsection{Anti-adiabatic limit $\omega_0\rightarrow\infty$}

For $\omega_0>0$, we can integrate out the phonons. In the anti-adiabatic
limit $\omega_0\rightarrow \infty$, the electron-phonon coupling reduces to an
electronic interaction proportional to the square of the hopping
operator \cite{Assaad16,Goetz22},
\begin{equation}
\hat{H} = -  t  \sum_{b}   \hat{K}_b
	   -\left(\lambda + \frac{g^2}{2k}\right) \sum_{b}  \hat{K}_b^2 \,.
\end{equation}
In this work we only consider $\lambda>0$, such that the additional electronic 
interaction has the same sign as the effective interaction in the anti-adiabatic limit.
For $N=2$ (\ie, spin $\frac{1}{2}$), it can be rewritten in terms of the
generators of the $\mathrm{O}(4)$ symmetry,
\begin{equation}
\label{Kb2.eq}
- \frac{1}{4}  \hat{K}_b^2 =  \hat{\boldsymbol{S}}_{\i}\cdot \hat{\boldsymbol{S}}_{\j} + \hat{\boldsymbol{\eta }}_{\i}\cdot \hat{\boldsymbol{\eta}}_{\j}\,.
\end{equation}
For $\omega_0  \rightarrow \infty$ and $\frac{g^2}{2k}+\lambda>0$, the SSH electron-phonon interaction hence favors an
AFM/CDW/SC ground state \cite{Assaad16}, as has been verified for $\lambda=0$ \cite{Cai21,Goetz22}.

\subsubsection{Vanishing direct hopping ($t=0$)}\label{subsec:t0}

In the limit $t=0$, electronic hopping is phonon mediated and Hamiltonian~(\ref{eq:SSH}) simplifies to 
\begin{equation}\label{eq:SSH-t=0}
\hat{H}= \frac{g}{\sqrt{2 m \omega_0}}\sum_b \left( \hat{a}_b^{\dagger} +
  \hat{a}_b^{\phantom{\dagger}} \right) \hat{K}_b
+ \omega_0 \sum_b \hat{a}_b^{\dagger}  \hat{a}_b^{\phantom{\dagger}} - \lambda \sum_b \hat{K}_b^2 \,.
\end{equation}
Here, we expressed the phonons in second
quantization. Equation~(\ref{eq:SSH-t=0}) has an additional local $\mathbb{Z}_2$
symmetry, explicitly broken at any nonzero $t$, represented by the local {\it star operators}
\begin{equation}
\hat{Q}_{\i}  = \hat{p}_{\i}(-1)^{\sum_{\bm{\delta}} \hat{a}_{\langle \i,\i+ \bm{\delta} \rangle}^{\dagger} \hat{a}_{\langle \i,\i+ \bm{\delta} \rangle}^{\phantom{\dagger}}} 
\end{equation}
obeying
\begin{equation}
{[\hat{H},\hat{Q}_{\i}]} = 0, \quad  [\hat{Q}_{\i},\hat{Q}_{\j}] = 0, \quad \hat{Q}_{\i}^2 = \mathds{1}\,.
\end{equation}
$\hat{Q}_{\i}$ captures the fermion parity at site $\i$ and
the parity of the phonon excitations on the bonds $\langle
\i,\i+\bm{\delta}\rangle$ connected to site
$\i$. Here, $\bm{\delta}\in\{\pm\bm{a}_x,\pm\bm{a}_y\}$.  Because we do \textit{not} impose the Gauss law on the
states of the Hilbert space $\hat{Q}_{\i}|\cdot\rangle =  |\cdot\rangle$,
Eq.~(\ref{eq:SSH-t=0}) corresponds to an unconstrained $\mathbb{Z}_2$ gauge
theory coupled to fermions.

Bosons  and fermions acquire $\mathbb{Z}_2$  charge,
\begin{equation}
\left\{ \hat{Q}_{\i},  \hat{c}_{\i,\sigma}^{\dagger}  \right\}    =  \left\{ \hat{Q}_{\i},  \hat{a}_b^{\dagger}  \right\}   =0. 
\end{equation}
At $t=0$, $\hat{Q}_{\i}$ is a constant of motion, so that these particles cannot
propagate in space:
\begin{equation}\label{eq:local_green}
  \langle \hat{c}^{\dagger}_{\bm{i},\sigma} (\tau)
  \hat{c}^{\phantom\dagger}_{\bm{j},\sigma} (0)   \rangle = 
  \delta_{\bm{i},\bm{j}}   \langle \hat{c}^{\dagger}_{\bm{i},\sigma} (\tau)
  \hat{c}^{\phantom\dagger}_{\bm{i},\sigma} (0)   \rangle \, , 
\end{equation}
\begin{equation}\label{eq:local_boson}
  \langle \hat{a}^{\dagger}_{b} (\tau)
  \hat{a}^{\phantom\dagger}_{b'} (0)   \rangle = 
  \delta_{b,b'}   \langle \hat{a}^{\dagger}_{b} (\tau)
  \hat{a}^{\phantom\dagger}_{b} (0)   \rangle \, . 
\end{equation}

\begin{figure}[t]
  \includegraphics[width=0.7\linewidth]{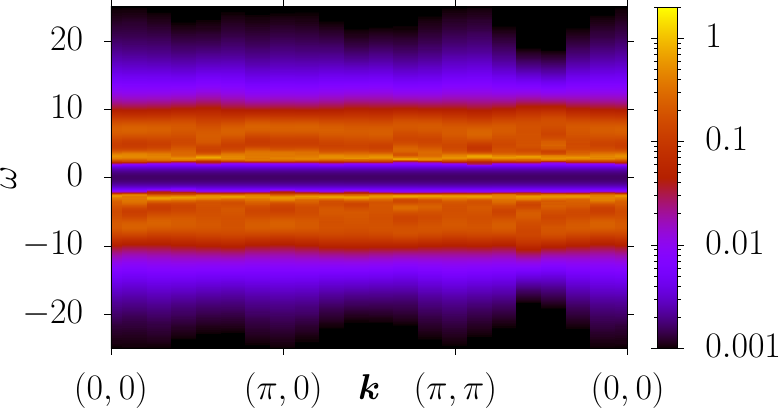}
  \caption{\label{orthogonal.fig} Single-particle spectral function $A(\bm{k},\omega)$ of the
    $c$-fermions at $t=0$. Here, $\beta = L = 14$,  $\omega_0=2.0$, $\tilde{g}=\sqrt{2/k}g=2$, $\lambda=0.5$.   }
\end{figure}

In Fig.~\ref{orthogonal.fig}, we show the single-particle spectral function of
the  $c$ fermions at  $t=0$, revealing a gap and the absence of dispersion.
The single-particle gap corresponds  to   the  energy  difference  between    the low-lying  $\hat{Q}_i $  sectors.
  
To capture the  physics  in this  limit,  we fractionalize  the $c$ fermion
into  a $\mathbb{Z}_2$ matter  field and an $f$ fermion~\cite{Ruegg10},
\begin{equation}\label{eq:fractionalize}
\hat{c}^{\dagger}_{\i,\sigma}   =   \hat{\tau}^z_{\i} \hat{f}^{\dagger}_{\i,\sigma}  
\end{equation}    
with the constraint
\begin{equation}
  \hat{\tau}^{x}_{\i}    \hat{p}_{\i}  = 1 
\end{equation}
and Pauli matrices $\hat{\tau}^{x}_{\i} $ and  $\hat{\tau}^{z}_{\i}$. The  constraint implies   that
the  $\mathbb{Z}_2$  matter  field,  $\hat{\tau}^{z}_{\i} $,    carries  the
$\mathbb{Z}_2$    charge    and  $ \hat{f}^{\dagger}_{\i,\sigma}  $
the quantum numbers  of the electron, namely, its global  U(1) charge and spin.  
This is in contrast to Refs.~\cite{PhysRevB.72.205124,PhysRevLett.102.065301},
where a gauge invariant (\ie, no $\mathbb{Z}_2$ charge)
$c$ fermion is replaced by the product of a $\mathbb{Z}_2$-charged $f$ fermion
and a slave spin.

In this representation, the Hamiltonian takes the form
\begin{eqnarray}\label{eq:SSH-t=0_f}\nonumber
\hat{H} &=&  \frac{g}{\sqrt{2 m \omega_0}}\sum_{b} \left( \hat{b}_b^{\dagger} +
  \hat{b}_b^{\phantom{\dagger}} \right) \hat{K}^{f}_b  
  + \omega_0 \sum_b \hat{b}_b^{\dagger}  \hat{b}_b^{\phantom{\dagger}}\\
  &&\quad- \lambda \sum_b \left( \hat{K}_b^f\right)^2,
\end{eqnarray}
with 
\begin{equation}
	\hat{K}^f_b =   \sum_{\sigma }
	  \left( \hat{f}^{\dagger}_{\i,\sigma}
            \hat{f}^{\phantom{\dagger}}_{\j,\sigma}    + \text{h.c.} \right),  \, \,  \hat{b}_b^{\dagger}    =  \hat{\tau}^z_{\i} \hat{a}_b^{\dagger}   \hat{\tau}^z_{\j} 
\end{equation} 
for  $b =  \left\langle \i,\j \right\rangle$. The constraints can be written as
\begin{equation}
\hat{Q}_{\i}  = \hat{\tau}^x_{\i}(-1)^{\sum_{\bm{\delta}} \hat{b}_{\langle \i,\i+ \bm{\delta} \rangle}^{\dagger} \hat{b}_{\langle \i,\i+ \bm{\delta} \rangle}^{\phantom{\dagger}}}\,. 
\end{equation}
Because $\hat{b}_b^{\dagger} $ and $\hat{f}^{\dagger}_{\i,\sigma}$ carry  no
$\mathbb{Z}_2$ charge,  they  can  propagate. In particular,
$\hat{b}^{\dagger}$  can  condense. The orthogonal fermion representation is a
good starting point for mean-field  theories.

\subsubsection{Adiabatic limit $\omega_0 \rightarrow 0 $ at $t=0$}

To discuss this limit, it is convenient to return to first quantization,
\begin{equation}
  \hat{q}_b =   
  \frac{1}{\sqrt{2 m \omega_0}} \left( \hat{b}_b^{\dagger} +
    \hat{b}_b^{\phantom{\dagger}}  \right)\,.
\end{equation}
For  $\omega_0 \rightarrow 0$, Hamiltonian~(\ref{eq:SSH-t=0_f}) becomes
\begin{equation}\label{eq:omega0t0}
  \hat{H}=   g\sum_{b}   \hat{q}_b  \hat{K}^{f}_b  
  +   \frac{k}{2} \sum_b  \hat{q}_b^2 - \lambda \sum_b \left( \hat{K}_b^f\right)^2\,.
\end{equation}
Equation~(\ref{eq:omega0t0}) has reflection positivity (see Ref.~\cite{Lieb94}) for any
line P parallel to the lattice vectors and cutting through the center of the
bonds.  Thereby, the flux through a circuit with lattice sites corresponding to
the corners of a plaquette will take the value $\pi$ \cite{Lieb94}.  A circuit
is a set of lattice sites $i_1, i_2, \dots, i_n, i_1$ with
$q_{i_m,i_{m+1}} \neq 0 \, \, \forall \, \,m $.  Hence, if all plaquettes turn out to
define circuits, Lieb's theorem \cite{Lieb94} implies that the $f$ fermions acquire a Dirac
spectrum.

\subsubsection{Anti-adiabatic limit $\omega_0 \rightarrow\infty$ at $t=0$}
For $\omega_0\rightarrow\infty$ and $t=0$, Eq.~(\ref{eq:SSH}) becomes equivalent
to the model studied in Ref.~\cite{Assaad16}, describing fermions coupled to
quantum Ising spins. This can be seen by approximating the phonons as hard-core
bosons with the constraint $(\hat{a}_b^{\dagger})^2=0$ and defining Ising variables
\begin{equation}
\hat{s}^x_b = 2\hat{a}_b^{\dagger} \hat{a}_b^{\phantom{\dagger}}-1, \quad \hat{s}^z_b = \hat{a}_b^{\dagger} + \hat{a}_b^{\phantom{\dagger}} \,.
\end{equation}

\subsection{Previous work}\label{sec:previous-work}

Until recently, most investigations of the 2D SSH model~(\ref{eq:SSH_generic})
were based on mean-field treatments or the assumption of classical phonons
($\omega_0=0$). The focus was on the true VBS pattern in the ground state
\cite{Mazumdar87,PhysRevB.37.9546,PhysRevB.39.12324,PhysRevB.39.12327}, the possible existence of a 
multi-mode Peierls state with no well-defined ordering wave vector
\cite{JPSJ.69.1769,JPSJ.72.1995,JPSJ.73.2473}, and the emergence of AFM order
from additional electron-electron repulsion \cite{PhysRevB.37.9546,PhysRevB.46.1710,YuanKopp2001,PhysRevB.65.085102,JPSJ.73.2777}.
These works were followed by unbiased QMC investigations on the honeycomb
lattice \cite{pub.1051655427}, the Lieb lattice \cite{li2020quantum}, and the
square lattice considered here \cite{Xing21,Cai21,Goetz22,cai2022robustness,Feng22}.
On the latter, a unique VBS ground state with ordering wave vector $(\pi,\pi)$, suggested by
mean-field theory, is well established. Surprisingly, the SSH model also
supports AFM order from electron-phonon coupling at sufficiently high phonon
frequencies \cite{Cai21,Goetz22}. 

\section{Method}\label{sec:method}

We simulated the modified SSH model~(\ref{eq:SSH}) using an auxiliary-field QMC
approach. To decouple the electron-phonon interaction, we first rewrite the
Hamiltonian to make the interaction term a perfect square,
\begin{eqnarray}
  \hat{H} &=&  \hat{H}_{t} + \hat{H}_{\lambda} + \hat{H}_{\mathrm{ph}}\,, \\
  \hat{H}_{t} &=&  - t \sum_b  \hat{K}_b \,, \quad \hat{H}_{\lambda} =  - \lambda\sum_b \left( \hat{K}_b - \frac{g}{2\lambda} \hat{X}_b  \right)^2 \,,  \nonumber \\
  \hat{H}_{\mathrm{ph}} &=&   \sum_b \frac{1}{2m} \hat{P}_b^2 +  \left(\frac{k}{2} + \frac{g^2}{4 \lambda} \right) \hat{X}_b^2 \nonumber \, . 
\end{eqnarray}
Using a Trotter decomposition with step size $\Delta \tau=\beta/L_\tau$, the
partition function reads  as
\begin{eqnarray}\label{eq:symmZ}
  Z &=& \Tr{e^{-\beta \hat{H}}}  = \Tr{ \left [ \left( e^{-\Delta\tau \hat{H}}  \right) ^{L_{\tau}}\right] } \nonumber \\
  e^{-\Delta \tau \hat{H} } &=&  e^{-\frac{\Delta\tau}{2}\hat{H}_t} 
                                \left(  \prod_{b=1}^{N_\text{b}} e^{-\frac{\Delta\tau}{2} \hat{H}_{\lambda,b}}  \right)
                                e^{-\Delta\tau \hat{H}_{\mathrm{ph}}} \\
  &&    \times  \left(  \prod_{b=N_\text{b}}^{1} e^{-\frac{\Delta\tau}{2} \hat{H}_{\lambda,b}}  \right)
       e^{-\frac{\Delta\tau}{2}\hat{H}_t}  + \mathcal{O}(\Delta\tau^{3}) \, 
       \nonumber
\end{eqnarray}
with $N_\text{b}=2N_\text{s}$ the total number of bonds and $\beta=1/T$ the inverse temperature. To preserve the
Hermiticity of the Hamiltonian, we used a symmetric Trotter decomposition
\cite{Hirsch82,Raedt83}. In Appendix~\ref{sec:appendix_trotter}, we compare this
decomposition with an asymmetric one that breaks the Hermiticity. Electrons and
phonons can now be decoupled with a discrete Hubbard-Stratonovich
transformation \cite{Hirsch83,Assaad97,Imada97,ALF22},
\begin{eqnarray}
&&  e^{\frac{\lambda \Delta \tau}{2} \left(\hat{K}_b-\frac{g}{2 \lambda}\hat{X}_b \right)^2}   \\
&& =  \frac{1}{4} \sum_{l=\pm1,\pm2} \gamma(l) e^{\sqrt{\frac{\Delta\tau\lambda}{2}}\eta(l)\left(\hat{K}_b-\frac{g}{2 \lambda}\hat{X}_b \right)} 
    +  \mathcal{O}[(\Delta\tau \lambda)^{4}] \, , \nonumber
\end{eqnarray}
where
\begin{eqnarray}
  \gamma(\pm1)= 1+\sqrt{6}/3 \,, \quad \eta(\pm1)=\pm\sqrt{2\left(3-\sqrt{6}\right)}\,, \\
  \gamma(\pm2)=1-\sqrt{6}/3 \,, \quad \eta(\pm2)= \pm\sqrt{2\left(3+\sqrt{6}\right)}\,. \nonumber
\end{eqnarray}
Since the Trotter decomposition introduces a systematic error of order
$\Delta \tau^3$ \cite{Fye86}, we can assume the Hubbard-Stratonovich transformation to be exact.
To evaluate the trace over the phonons in the partition function we use a path
integral in real-space representation with the eigenstates $|x_b \rangle $ and
eigenvalues $\x_b$ of the position operator,
\begin{eqnarray}
  Z &=&  \sum_{\left\{l_{b,\tau,\alpha}\right\}} \left(\prod_{b,\tau,\alpha} \frac{ \gamma(l_{b,\tau,\alpha})}{4} \right)  \det\left[ 1+ B(\beta,0) \right] \\
    && \quad \times \int \prod_{b,\tau} d\x_{b,\tau} e^{ -\sum_b \bm{\x}_b^{\mathrm{T}} A \bm{\x}_b - \sum_b  \v_b^{\mathrm{T}} \bm{\x}_b } \,,  \nonumber 
\end{eqnarray}
where we used
\begin{eqnarray}\label{eq:A_v}
    \bm{\x}_b^{\mathrm{T}} &=& \left( \x_{b,1}, \x_{b,2},\dots, \x_{b,N_\tau} \right) \,, \\
    v_{b,\tau} &= &g \sqrt{\frac{\Delta \tau}{8 \lambda}}  \left(\eta(l_{b,\tau,2})+\eta(l_{b,\tau-1,1}) \right) \,, \nonumber\\
    A_{k,l} &=&\Delta\tau\left( \alpha \delta_{k,l} - \gamma \left(\delta_{k,l+1} + \delta_{k,l-1} \right) \right)\,, \nonumber \\
    \alpha &=& \frac{k}{2}+ \frac{g^2}{4\lambda}+\frac{m}{\Delta \tau^2} \,,
    \quad \gamma = \frac{m}{2 \Delta \tau^2}\,. \nonumber
\end{eqnarray}
The index $\alpha=1,2$ is needed as the symmetric Trotter decomposition
produces two Hubbard-Stratonovich decompositions per time slice and bond. For
the electronic part, we rewrite the trace as a determinant \cite{Blankenbecler81},
\begin{eqnarray}\label{eq:B}
  B(\tau_1,\tau_2) &=& \prod_{\tau=\tau_2+\Delta \tau}^{\tau_1} e^{\frac{\Delta \tau}{2}  t \sum_b  K_b} \left(\prod_{b=1}^{N_\text{b}} e^{\sqrt{\frac{\Delta\tau\lambda}{2}}\eta(l_{b,\tau,1})K_b} \right) \nonumber \\
                  &  &\times \left(\prod_{b=N_\text{b}}^{1} e^{\sqrt{\frac{\Delta\tau\lambda}{2}}\eta(l_{b,\tau,2})K_b} \right) e^{\frac{\Delta \tau}{2}  t \sum_b  K_b} \,.
\end{eqnarray}

We can interpret the matrix $A$ as a tight-binding Hamiltonian on a periodic
chain with $L_\tau$ sites, hopping $\alpha$, and on-site potential
$\gamma$. The eigenvalues of $A$ are
\begin{eqnarray}
a_n &=& \Delta\tau \left( \alpha - 2\gamma \cos \nu_n  \right)  \\
&=&  \Delta\tau \left[\frac{k}{2} + \frac{g^2}{4 \lambda} + \frac{m}{\Delta \tau^2}(1-\cos \nu_n)\right]\,, \nonumber
\end{eqnarray}
with $\nu_n=\frac{2\pi}{L_\tau}n$ and $1\leq n\leq L_\tau$. If $k + \frac{g^2}{4 \lambda} \leq0$, 
the eigenvalues of $A$ can become zero or negative and the integral over the phonons does not converge.
But for $\lambda>0$ the matrix $A$ is positive definite 
for the whole parameter range and we can integrate out the
phonons to obtain 
\begin{eqnarray}\label{eq:Z}
  Z & \propto &
              \sum_{\left\{l_{b,\tau,\alpha}\right\}} \left(\prod_{b,\tau,\alpha} \gamma(l_{b,\tau,\alpha}) \right) \\
           & &\quad\times
              \exp{\left\{ \frac{1}{4} \sum_b \v_b^{\mathrm{T}} A^{-1} \v_b \right\} }  \det\left[ 1+ B(\beta,0) \right] \, . \nonumber
\end{eqnarray}

The summation over the auxiliary fields $\{l_{b,\tau,\alpha}\}$ is done stochastically
with QMC methods employing single-spin-flip and global updates, which are
accepted according to the Metropolis-Hastings algorithm
\cite{Metropolis53,Hastings70}.  For the global updates, we randomly choose a
rectangular section of auxiliary fields in the (2+1)D configuration space and
swap it with a rectangle of the same size but displaced by $L_{\tau}/2$ in
the imaginary-time direction.

Additionally, we used a $\beta$-doubling method to reduce warm-up times.
We started by running a simulation with a given parameter set for
some time with an inverse temperature $\beta_1$ smaller than the final value
$\beta$. Then, we used the last configuration of this run as a starting
configuration for a simulation with a higher inverse temperature $\beta_2$, 
$\beta_1<\beta_2\leq 2 \beta_1$, identifying
$\eta(l_{b,\tau+\beta_1,\alpha})=\eta(l_{b,\tau,\alpha})$ at the
beginning. After two or three such steps, we reached the final inverse
temperature $\beta$.

Appendix~\ref{sec:appendix_comparison} contains a comparison of the method
described here and used for the results with two other methods that do not
involve integrating out the phonons.  For  the parameters considered, the  
present  approach   provides a  substantial  speedup. 

\section{Results}\label{sec:results}

For the simulations, we absorbed $k$ into a renormalization of the phonon fields,
$\tilde{x}_{b,\tau}=\sqrt{{k}/{2}}x_{b,\tau}$, and set $\tilde{g}=\sqrt{2/k}g=2$,
$\lambda=0.5$, as well as $\Delta\tau=0.05$.

To detect the various phases, we measured imaginary-time-displaced correlation
functions
\begin{equation}\label{eq:corr_func}
\left[ S_{O}(\q,\tau) \right]_{\mu,\nu} =  \left \langle
  \hat{O}_{\mu}(\bm{q},\tau) \hat{O}_{\nu}(-\bm{q})  \right \rangle -
\left\langle \hat{O}_{\mu}(\bm{q}) \right \rangle \left\langle
  \hat{O}_{\nu}(-\bm{q}) \right \rangle
\end{equation}
and corresponding susceptibilities,
\begin{equation}
 \chi_{O}(\bm{q})  = \int_0^{\beta} d\tau  \Tr S_{O}(\q,\tau)\,,
\end{equation}
for several observables $\hat{O}$. The notation is general enough for correlators with
a matrix structure; the scalar case is obtained by dropping the
indices $\mu,\nu$. Here, $\bm{q}$ is a wave vector inside the first Brillouin
zone.

Because of the O(4) symmetry of Hamiltonian~(\ref{eq:SSH}), the spin correlator
is degenerate with charge and $s$-wave superconducting correlation functions (see
Sec.~\ref{sec:symm}). Therefore, we focus on the $z$ component of spin,
\begin{equation}
\hat{S}^z(\bm{q}) = \frac{1}{\sqrt{N_\text{s}}} \sum_{\i} e^{i \bm{q}\cdot \i}  \left(\hat{n}_{\i,\uparrow} - \hat{n}_{\i,\downarrow} \right) \, .
\end{equation}
To detect the breaking of particle-hole symmetry, we measured correlation
functions of the parity $\hat{p}_{\bm{i}}$ [Eq.~(\ref{eq:par})]. Additionally, we
calculated dimer correlations to detect VBS order that breaks the discrete
$C_4$ symmetry of the square lattice,
\begin{eqnarray}\label{eq:dimer_order}
\hat{\Delta}_{\mu}(\bm{q})  &=& \frac{1}{\sqrt{N_\text{s}}} \sum_{\i} e^{i \bm{q}\cdot \i} \hat{\Delta}_{\i,\mu} \, , \\
\hat{\Delta}_{\i,\mu} & = & \hat{S}_{\sigma,\rho}(\i) \hat{S}_{\rho,\sigma}(\i+\bm{a}_{\mu}) \,, \nonumber
\end{eqnarray}
where 
\begin{equation}
\hat{S}_{\sigma,\rho}(\i) = \hat{c}^{\dagger}_{\i,\sigma} \hat{c}^{\phantom{\dagger}}_{\i,\rho} - \frac{1}{2} \delta_{\sigma,\rho} \,. 
\end{equation}
The vectors $\bm{a}_{x}^{\mathrm{T}}=(1,0)$ and $\bm{a}_{y}^{\mathrm{T}}=(0,1)$ connect site $\i$ to its nearest neighbors and $\mu\in\{x,y\}$.

We further present results for the correlation ratios
\begin{equation}
R_{S,O} = 1- \frac{\Tr S_{O}(\bm{q}_{O}+\Delta\bm{q})}{\Tr S_{O}(\bm{q}_{O})}\,,
\end{equation}
where $|\Delta\bm{q}|=2\pi/L$ is the shortest wave vector on an $L\times L$ lattice and
$\bm{q}_{O}$ the ordering wave vector of observable $O$. The correlation ratio
is a renormalization group invariant quantity and takes on values close to zero/one in the
disordered/ordered phase. For the susceptibilities, correlation ratios can be
defined analogously.

The single-particle spectral function $A(\bm{k},\omega)$,
accessible in angular-resolved photoemission, was calculated from the imaginary-time Greens
function with the stochastic maximum entropy method \cite{Sandvik98,Beach04a} via
\begin{equation}
  \langle \hat{c}^{\phantom\dagger}_{\bm{k},\sigma} (\tau)
  \hat{c}^{\dagger}_{\bm{k},\sigma} (0)   \rangle   =  \frac{1}{\pi} \int
  \text{d} \omega  \, \frac{e^{-\tau \omega} }  {  1 + e^{-\beta\omega} }
  A(\bm{k}, \omega) \,.
\end{equation}
The dynamical structure factors for spins and dimers were computed from
\begin{equation}
\Tr S_{O}(\bm{q},\omega) = \frac{\Tr \chi''_{O}(\bm{q},\omega)}{1-e^{-\beta\omega}} \, ,
\end{equation} 
where the imaginary part of the dynamical susceptibility was obtained by inverting
 \begin{equation}
\Tr S_{O}(\bm{q},\tau) = \frac{1}{\pi} \int d\omega \frac{e^{-\tau\omega}}{1-e^{-\beta\omega}} \Tr \chi''_{O}(\bm{q},\omega)
\end{equation}
with the maximum entropy method \cite{ALF22}.

Finally, we also measured observables that depend on phonon variables, such as the flux operator $\hat{\Phi}_{\i}$. 
The latter is defined as the product over the effective hoppings on the bonds $b$
of an elementary plaquette $\Box_{\bm{i}}$,
\begin{equation}\label{eq:flux}
\hat{\Phi}_{\bm{i}}=\prod_{b\in\Box_{\bm{i}}}\left(-t+g\hat{X}_b\right) \,,
\end{equation}
where $\bm{i}$ is one of the four corner sites of a plaquette. Because the
phonons were integrated out, such observables are not directly
accessible but require a source term in the action, as discussed in
Appendix~\ref{sec:appendix_phonon}.

Throughout   the  paper,  we  have opted for  a  $\beta = L$ scaling.  This scaling captures  ground-state properties  in Lorentz-symmetric  phases,  such as the AFM  and  $(\pi,0)$ VBS, and  critical points. 
As  we  will see  below,  both the AFM and $(\pi,0)$ VBS phases correspond to mass terms  of  emergent  Dirac  fermions. These mass-generating  symmetry-breaking fields  do not  break Lorentz symmetry. In contrast, phases  that are  not  characterized by  Lorentz  symmetry will be dominated by  finite-temperature  effects.  

\subsection{Dynamically generated $\pi$ flux} \label{sec:flux}

\begin{figure}[t]
  \includegraphics[width=0.7\linewidth]{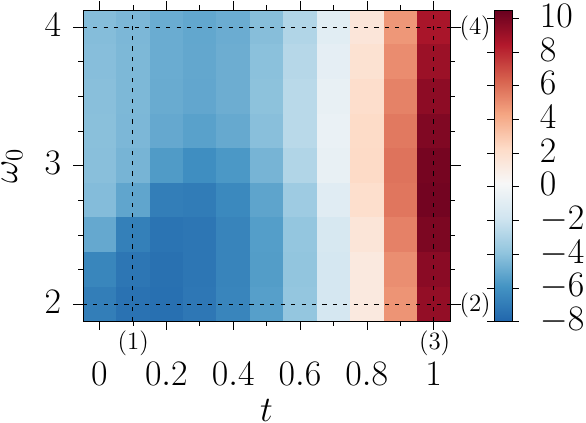}
  \caption{\label{fig:flux}Average flux per plaquette
    $\frac{1}{N_\text{s}}\sum_{\bm{i}}\langle \hat{\Phi}_{\bm{i}} \rangle$ for 
    $\beta=L=8$. See also Figs.~\ref{fig:scal_t01}(a) and \ref{fig:scal_t10}(a) for the flux  along the dotted lines (1) and (3)
    at fixed $t$ and varying phonon frequency and Figs.~\ref{fig:scal_w20}(a) and \ref{fig:scal_w40}(a)
    for the flux along the dotted lines (2) and (4) at fixed $\omega_0$ and varying hopping.}
\end{figure}

We first consider the flux per elementary plaquette [Eq.~(\ref{eq:flux})] in the
$t$-$\omega_0$ plane, shown in Fig.~\ref{fig:flux}. In an electron-phonon context,
$g\hat{X}_b$ in Eq.~(\ref{eq:SSH}) should be a small perturbation to the bare hopping $t$.  Hence, in
this regime, the flux per plaquette is positive. The
$\hat{c}^{\dagger}_{\i,\sigma} $ operators create electrons and the hopping
matrix element leads to a $(\pi,\pi)$-nested Fermi surface. The latter gives
rise to instabilities at $\bm{q}=\bm{M}$. Possible orderings include an AFM
phase or a $(\pi,\pi)$ VBS phase, as observed for $\lambda=0$
\cite{Xing21,Cai21,Goetz22,cai2022robustness,Feng22}.

In the opposite, phonon-assisted hopping limit (\ie, at $t=0$), our model
reduces to an unconstrained $\mathbb{Z}_2$ gauge theory, see
Sec.~\ref{subsec:t0}.  The $c$ fermions acquire a locally conserved
$\mathbb{Z}_2 $ charge and cannot propagate in space, as demonstrated in
Fig.~\ref{orthogonal.fig}.  However, gauge invariant quantities such as the
local spin or local charge can propagate.  Hence, the $c$ fermion corresponds to
a so-called orthogonal fermion \cite{Nandkishore12}.  To understand the
single-particle physics, we have to adopt the $f$ fermions
[Eq.~\ref{eq:fractionalize})]. The latter carry no locally conserved
$\mathbb{Z}_2$ charge, can hop from site to site, and acquire a phase of $0$ or
$\pi$ when circulating around a plaquette.  A phase of $\pi$ is favored by
Lieb's theorem \cite{Lieb94} and confirmed by the numerical results in
Fig.~\ref{fig:flux}.

The generated $\pi$-fluxes cause the $f$ fermions to acquire a Dirac dispersion
relation \footnote{The  $\pi$-flux  lattice  as  discussed e.g. in  Ref.~\cite{Toldin14}  is  a possible  
lattice  regularization of the  Dirac  equation, },
which has important
consequences for the understanding of the phase diagram.  In particular, we can
classify the interaction-generated ordered states in terms of mass terms that
generate a single-particle gap and break discrete or continuous symmetries \cite{Ryu09}.
Of particular importance are the two $(\pi,0)$ VBS masses and the three
AFM mass terms, which break the $\mathrm{C}_4$ lattice symmetry
and the $\mathrm{SU}(2)$ spin symmetry, respectively.  The
Dirac vacuum allows for topological terms in the action, which play a key role
in the understanding of DQCPs \cite{Abanov00,Tanaka05,Senthil06}.
    
Our symmetry arguments are valid only at $t=0$.  Beyond this limit,
$\hat{Q}_{\i}$ is not a good quantum number and the $\mathbb{Z}_2$ charge
is not locally conserved.  As a consequence, the $c$ fermions acquire a
dispersion, albeit small for small $t$. This is visible from the
single-particle spectral functions in Figs.~\ref{fig:Spectral_t01}(Ia)-(Ie), 
that will be discussed in more detail in the following section.
The lack of dispersion of the $c$ fermions for small values of $t$ suggests that
it is still appropriate to work in the $f$ basis.  The hopping of the
$f$ fermions reads as
\begin{equation}
  \sum_{b =\langle\i,\j\rangle}   \hat{\tau}^{z}_i \left(    -t   +  g  \hat{X}_{b} \right)  \hat{\tau}^{z}_j  \hat{K}^{f}_b. 
\end{equation}
Since $ \left( \hat{\tau}^z_i \right)^2 = 1$, the flux $\hat{\Phi}_{\bm{i}}$,
[Eq.~(\ref{eq:flux})] shown in Fig.~\ref{fig:flux} indeed
corresponds to the flux acting on the $f$ fermions. In Sec.~\ref{sec:low_w},
we  will further argue that  the  free  energy  is  an analytical
function of  $t$  at $t=0$, so  that  the $t=0$  physics  is  adiabatically
connected to a region  around $t=0$ set by the convergence  radius  of  the  series.

As a function of $t$, Fig.~\ref{fig:flux} reveals a crossover where the flux changes sign.  To
a first approximation, this sign change does not depend on $\omega_0$.  As we
will see below, it marks the crossover between a regime that can be understood in
terms of the $f$ fermions with a Dirac dispersion and a regime of $c$ fermions
with a nested Fermi surface.

\subsection{Deconfined quantum critical point}

\begin{figure}[t]
 \includegraphics[width=1\linewidth]{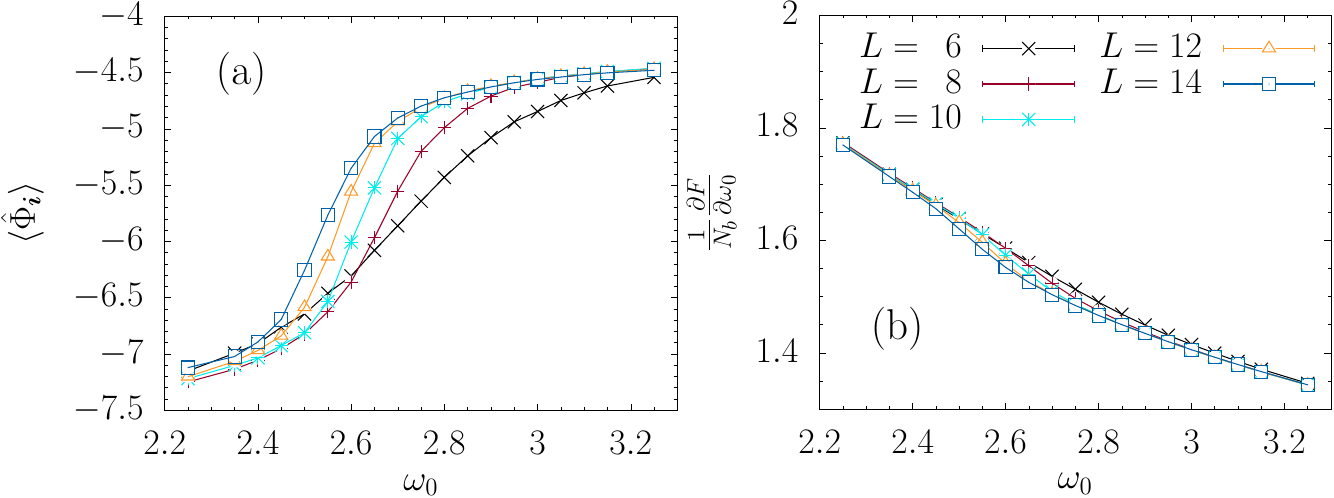}
  \caption{\label{fig:scal_t01} Flux and normalized free-energy derivative with
    respect to $\omega_0$. Here, $t=0.1$, $\beta=L$. See also Fig.~\ref{fig:df_dw}(a) in erratum.}
\end{figure}

Next, we study a cut along the frequency axis at a small but finite $t=0.1$
that explicitly breaks the local $\mathbb{Z}_2$ symmetry. Figure~\ref{fig:scal_t01}(a) 
shows the average flux per plaquette as a function of system size and phonon
frequency. The flux stays negative irrespective of $\omega_0$, corresponding to
a dynamically generated $\pi$ flux in each plaquette. The choice $\beta=L$
in the finite-size scaling is motivated by the Dirac band structure of the
$f$ fermions, as discussed below. The pronounced change of the flux 
around $\omega_0\approx2.6$ 
 suggests the possibility of a phase transition.

Another observable that carries information about a possible phase transition is
the free energy $F$. At a fixed electronic hopping $t$, its first derivative
with respect to $\omega_0$ is given by
\begin{equation}\label{eq:df_dw}
  \frac{\partial F}{\partial \omega_0} = m\omega_0 \sum_b \langle \hat{X}_b^2 \rangle - \frac{g}{2\omega_0}\sum_b\langle \hat{K}_b \hat{X}_b\rangle \,.
\end{equation}
Figure~\ref{fig:scal_t01}(b) reports results for this quantity as a function of
$\omega_0$ and for different $L$. The data reveal no jumps or kinks on the scale
considered, essentially ruling out a \textit{strongly} first-order transition.

\begin{figure}[tb]
 \includegraphics[width=1\linewidth]{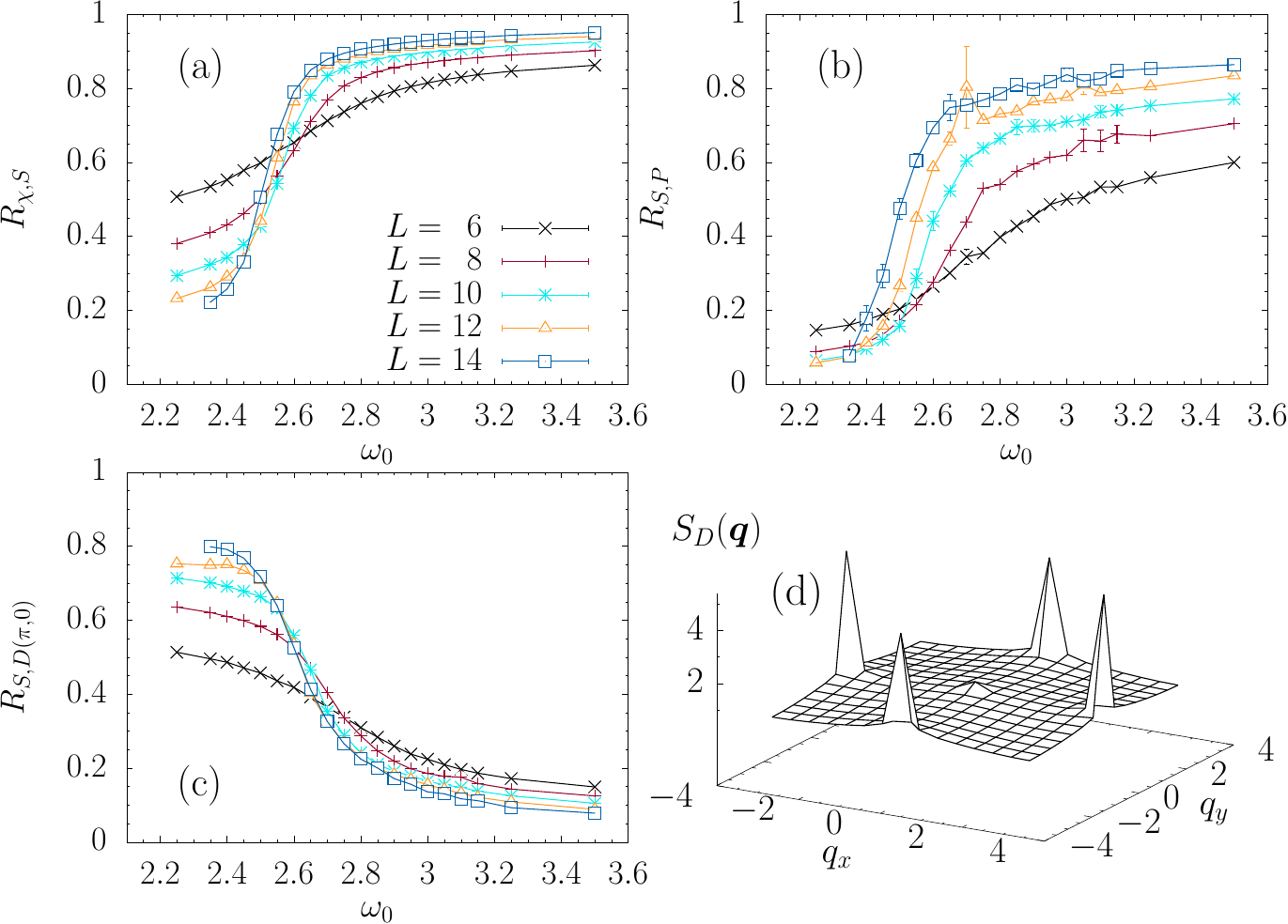}
  \caption{\label{fig:corr_t01}Correlation ratio for (a) spin susceptibility
    at $\bm{M}$, (b) parity correlation function at $\bm{\Gamma}$, and (c) dimer correlation function at $\bm{X}$ as a
    function of $\omega_0$ for $\beta=L$. (d) Equal-time dimer
    correlation function in the first Brillouin zone for  $\beta=L=14$, $\omega_0=2.35$.
    Here, $t=0.1$. }
\end{figure}

\begin{figure*}[ht]
    \includegraphics[width=1\linewidth]{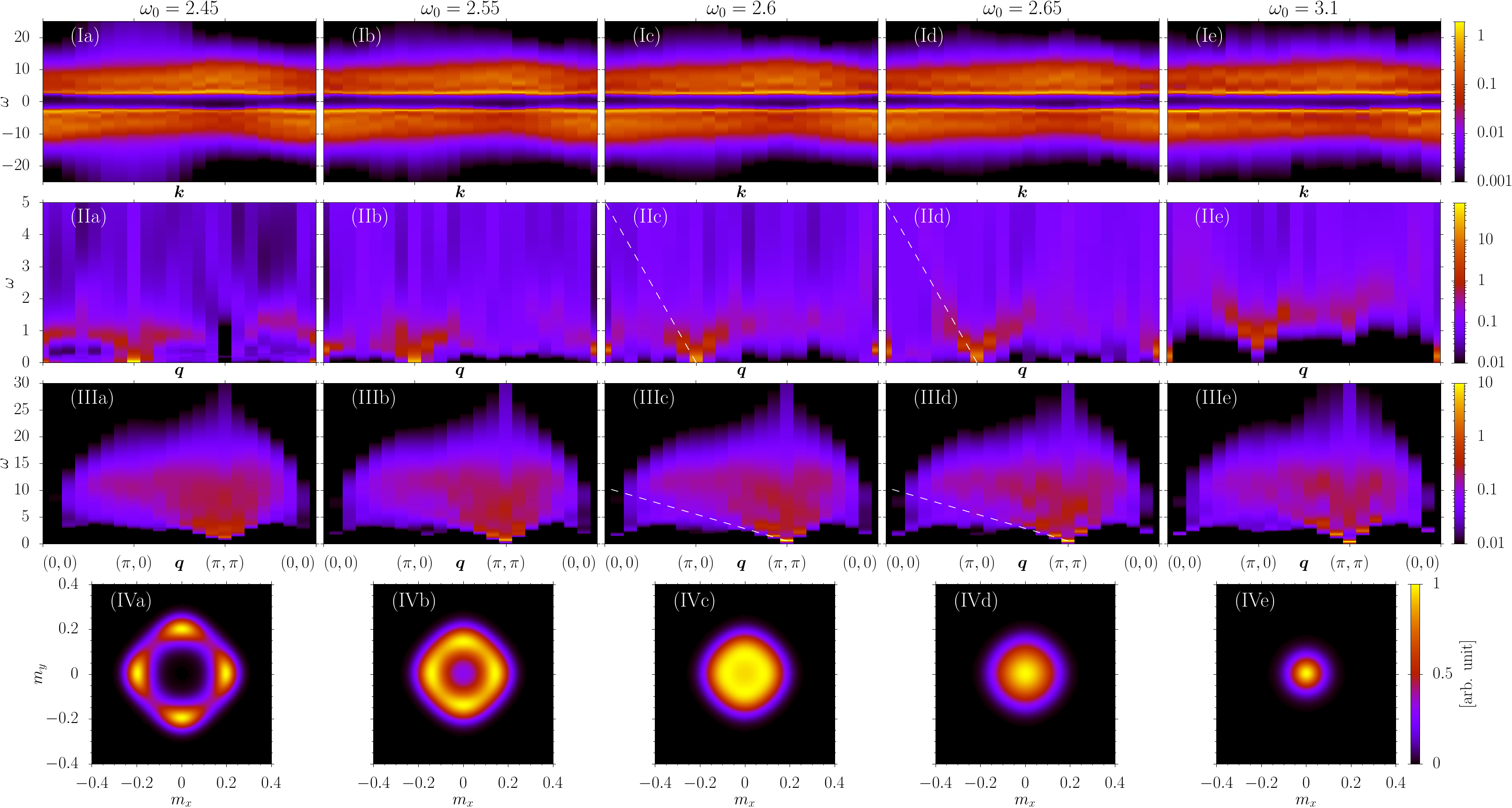} 
  \caption{\label{fig:Spectral_t01}(Ia)--(Ie) Single-particle spectral function
    $A(\bm{k}, \omega)$, (IIa)--(IIe) dynamical VBS structure factor
    $S_D(\bm{q},\omega)$, (IIIa)--(IIIe) dynamical spin structure factor
    $S_S(\bm{q},\omega)$, and (IVa)--(IVe) histogram of the VBS order parameter 
    $m_x$ and $m_y$  for different $\omega_0$ and $t=0.1$,
    $\beta=L=14$. White dashed lines in panels (IIc), (IId), (IIIc), and (IIId) are guides to the eye.}
\end{figure*}

In Figs.~\ref{fig:corr_t01}(a)--\ref{fig:corr_t01}(c), we present the correlation ratios of the spin 
 susceptibility as well as the equal-time parity and dimer correlation functions.
The correlation ratio scales as
\begin{equation}
  R_{\chi,O}=f\left(L^z/\beta,[\omega_0-\omega_0^c]L^{1/\nu}\right)\,,
\end{equation}
with the dynamical exponent $z$ and the correlation length exponent $\nu$. Here,
we neglect  corrections to scaling that will cause a meandering of
the crossing points with increasing $L$.  Unless indicated otherwise, we used
$\beta=L$ when measuring correlation ratios. This choice appears to be justified
by the dynamical dimer structure factor [Fig.~\ref{fig:Spectral_t01}(IIc)] and the
dynamical spin structure factor [Fig.~\ref{fig:Spectral_t01}(IIIc)] close to the
presumed critical point. Both quantities are consistent with a linear dispersion
relation around the ordering wave vector and hence with an exponent $z=1$.
Furthermore, we note that mass terms in the Dirac equation do not break Lorentz
symmetry so that this scaling remains justified even in the ordered phases.

The spin correlation ratio $R_{\chi,S}$ at wave vector $\bm{M}$
[Fig.~\ref{fig:corr_t01}(a)] reveals long-range AFM order at high phonon
frequencies. Simultaneously, the parity correlation ratio shows ordering at
$\bm{\Gamma}=(0,0)$.  By lowering $\omega_0$, AFM order disappears
but dimer correlations exhibit a marked increase. The equal-time dimer
correlation function is dominated by four peaks
at $\bm{q}=(\pi,0)$ and equivalent wave vectors [Fig.~\ref{fig:corr_t01}(d)]. The
corresponding correlation ratio at $\bm{X}$ increases with decreasing $\omega_0$
[Fig.~\ref{fig:corr_t01}(c)]. 

The correlation ratios are consistent with a phase transition from an AFM state
at high phonon frequencies to a $(\pi,0)$ VBS state in a  range $\omega_0^c  \simeq 2.4-2.6$.
Finite-size  effects make a more quantitative analysis
difficult. Regarding the nature of this phase transition,
several possible explanations exist.  We cannot exclude a weakly first-order
transition. The data are also consistent with an intermediate coexistence
region, especially since the quality of the results for the dimer correlation
ratio is limited by long autocorrelation times.  A third possibility
is a direct second-order transition. Because the AFM and VBS phases break
different symmetries, such a transition falls outside the Ginzburg-Landau paradigm
and is instead a candidate for a deconfined quantum critical point
\cite{Senthil04a,Senthil04b}.

To  obtain    better  insight in the nature  of  the phase  transition,  we  consider  spin, VBS, and $c$-fermion spectral functions.  Let us  start
with the theoretical expectations in the limit $t=0$. Because the VBS and spin order
parameters carry no $\mathbb{Z}_2$ charge, they will exhibit dispersive features
even for $t=0$. In contrast, the $c$ fermion has a $\mathbb{Z}_2$ charge and the
corresponding spectral function $A(\bm{k},\omega)=A(\omega)$,  as  shown in 
Fig.~\ref{orthogonal.fig}.
 At $t=0$, the
Hamiltonian is block diagonal in $\hat{Q}_{\i} $, which is a good quantum number.
Since $ \hat{c}^{\dagger}_{\i,\sigma} $ generates a $\mathbb{Z}_2$
charge, it causes changes between different $\hat{Q}_{\i} $ sectors. Hence, the
single-particle gap can be understood in terms
   of the energy difference between
different $\hat{Q}_{\i}$ sectors. States such as Dirac spin liquids,   or 
orthogonal  semi-metals   \cite{Nandkishore12, Assaad16, Gazit17}, would
exhibit gapless excitations in the spin sector, but gapped excitations in the
single-particle spectral function.  Hence, there is a priori no relation between the gaps
observed in the spin and the $c$-fermion sectors. 

Signatures of our theoretical expectations for $t=0$ are apparent in
Fig.~\ref{fig:Spectral_t01}, obtained for $t=0.1$. As for the case of  $t=0$ shown in 
Fig.~\ref{orthogonal.fig},   the spectral function is
essentially independent of $\bm{k}$ in Figs.~\ref{fig:Spectral_t01}(Ia)-(Ie).
Moreover, the dominant features show very little dependence on $\omega_0$, 
with substantial spectral weight at $\Delta_\text{sp} \simeq 2.5 $.    In the VBS phase
at $\omega_0 =2.45 $ [Fig.~\ref{fig:Spectral_t01}(IIIa)],  the dynamical spin
structure factor is reminiscent of gapped Dirac fermions due to the onset of VBS
order.  The spin gap can be read off as $\Delta_\text{s} \simeq 1.5 $. 
Since $t>0$ violates  the 
local $\mathbb{Z}_2$ symmetry, a  spin wave  can  decay  into  two
$c$ fermions. The  fact  that $\Delta_\text{s} < 2
\Delta_\text{sp}$ reflects  vertex corrections  accounting
for electron-hole  binding. 

For $\omega_0=2.45 $, we have VBS order that breaks the $C_4$
lattice symmetry. The  VBS spectral function in Fig.~\ref{fig:Spectral_t01}(IIa) shows a
sharp mode at $\omega = 0 $ and at the ordering wave vector $\bm{q} = (\pi,0) $
that accounts for the Bragg peak associated with this order.   In Fig.~\ref{fig:Spectral_t01}(IVa),  we  show  the  histogram 
of  the  VBS  order parameter   with $m_{\mu} = \hat{\Delta}_{\mu}(\bm{q}_{\mu})
/\sqrt{N_s}$,  $\mu=x,y$,  $\bm{q}_{x}=(\pi,0)$, and $\bm{q}_{y}=(0,\pi)$.
\footnote{We symmetrized the histograms by exploiting the $\mathrm{C}_4$ symmetry of the model 
and the arbitrariness of the minus sign in the definition of the order parameter
in Eq.~(\ref{eq:dimer_order}). }. We find four  peaks along  the  $x$ and  $y$
axes, reflecting the four-fold  degeneracy  of the $(\pi,0) $ VBS order parameter. 

\begin{figure}[b]
  \includegraphics[width=1\linewidth]{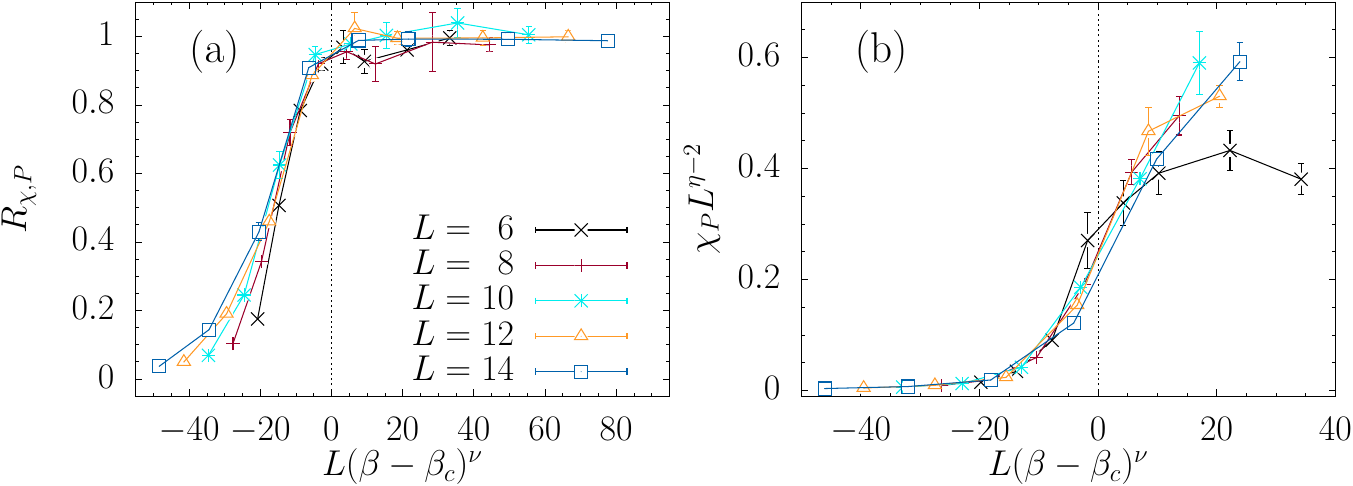} 
  \caption{\label{fig:par_t01}Data collapse of (a) the parity susceptibility
    correlation ratio and (b) the parity susceptibility using 2D Ising exponents
    ($\nu=1$, $\eta=1/4$). Here, $\bm{q}=\bm{\Gamma}$, $t=0.1$, $\omega_0=3.5$.  
   }
\end{figure}

It is beyond the scope of this work to  study  the  critical  exponents of the
purported DQCP. However, DQCPs have a number of hallmark signatures that can
be detected in the dynamical responses.  First, at criticality, the
$\mathrm{C}_4$ lattice symmetry is enlarged to U(1).  This symmetry enhancement
is captured by $S_D( \bm{q},\omega) $ in the form of a spectrum  with a linear mode.
Comparing Fig.~\ref{fig:Spectral_t01}(IIa) (deep in the VBS phase) to 
Fig.~\ref{fig:Spectral_t01}(IId) (close to the DQCP), we recognize that the Bragg
peak evolves towards a spectrum  with a linear dispersion relation.  
 Equivalently,  the  histograms   of  Fig.~\ref{fig:Spectral_t01}(IVa)-(IVd)  
reveal  that  the four-peak  structure    evolves  to  a  circle   upon approaching the critical point. 
 Similar    phenomena  have been  observed  in Ref.~\cite{Sandvik07}. 

Another DQCP hallmark is a single, continuous and direct
transition with emergent Lorentz symmetry. The corresponding theory has a single velocity.
Our data are consistent with this expectation: at criticality, the U(1)
velocity in Fig.~\ref{fig:Spectral_t01}(IIc)/(IId) compares favorably with the spin
velocity in Fig.~\ref{fig:Spectral_t01}(IIIc)/(IIId). Hence, several of the defining
properties of the DQCP are borne out by our results.

In the  AFM phase, Figs.~\ref{fig:Spectral_t01}(IIe) and~(IVe),  we  observe  a  gap in the  
dimer correlations,  a  Goldstone mode  in the  spin  correlations,  and  a
single central  peak in the  histogram.

In contrast to models of DQCPs with $\mathrm{SU}(2) \times \mathrm{C}_4$  \cite{Sandvik07} or
$\mathrm{SU}(2) \times \mathrm{U}(1)$ symmetry \cite{Liu18}, our model has an
$\mathrm{O}(4) \times \mathrm{C}_4$ symmetry.  In the AFM phase, the symmetry
is broken down to $\mathrm{U}(1) \times \mathrm{C}_4$.  As we will argue below,
this symmetry reduction occurs in two steps. The parity being a $\mathbb{Z}_2$
order parameter, we expect a finite temperature 2D Ising phase transition with
exponents $\nu=1$ and $\eta=1/4$.  These exponents yield a
satisfactory data collapse of the parity correlation ratio and susceptibility in
Fig.~\ref{fig:par_t01}. The collapse of the correlation ratio yields a critical
inverse temperature of $\beta_c  \simeq 4.5$, whereas the susceptibility data gives
$\beta_c \simeq 4.3$. Below the finite
temperature Ising transition, the symmetry is reduced to $\mathrm{SU}(2) \times
\mathrm{SU}(2)$ and either the odd or even parity sector is spontaneously
selected. Only at $T=0$ is the continuous SU(2) symmetry broken down to U(1).
 
\subsection{From assisted hopping to phonon-modulated direct hopping}\label{sec:low_w}

So far, we have focused on the small-$t$ regime of the phase
diagram, where the physics can be understood in terms of the
$f$ fermions with an underlying Dirac dispersion stemming from dynamically
generated $\pi$ fluxes. We now vary $t$, and thereby the ratio of direct to
phonon-assisted hopping, at a fixed phonon frequency $\omega_0=2.0$.

\begin{figure}[t]
 \includegraphics[width=1\linewidth]{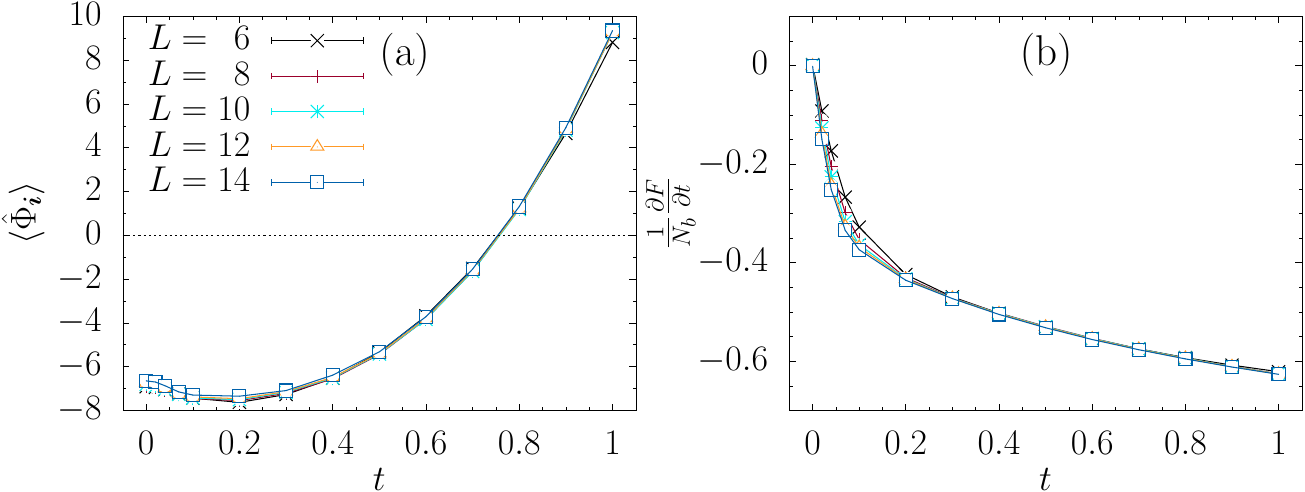}
  \caption{\label{fig:scal_w20}Flux and normalized free-energy derivative with
    respect to $t$. Here, $\omega_0=2.0$, $\beta=L$.}
\end{figure}

As previously revealed by Figs.~\ref{fig:phases} and \ref{fig:flux}, 
the flux changes its sign as a function of $t$. In Fig.~\ref{fig:scal_w20}(a),
we present results for the flux for different lattice sizes.  It varies
smoothly, with the sign changing at $t\approx0.75$.  Starting from
large values of $t$, the flux decreases until it reaches a minimum at $t\approx
0.2$ followed by a slight increase.

\begin{figure}[b]
 \includegraphics[width=1\linewidth]{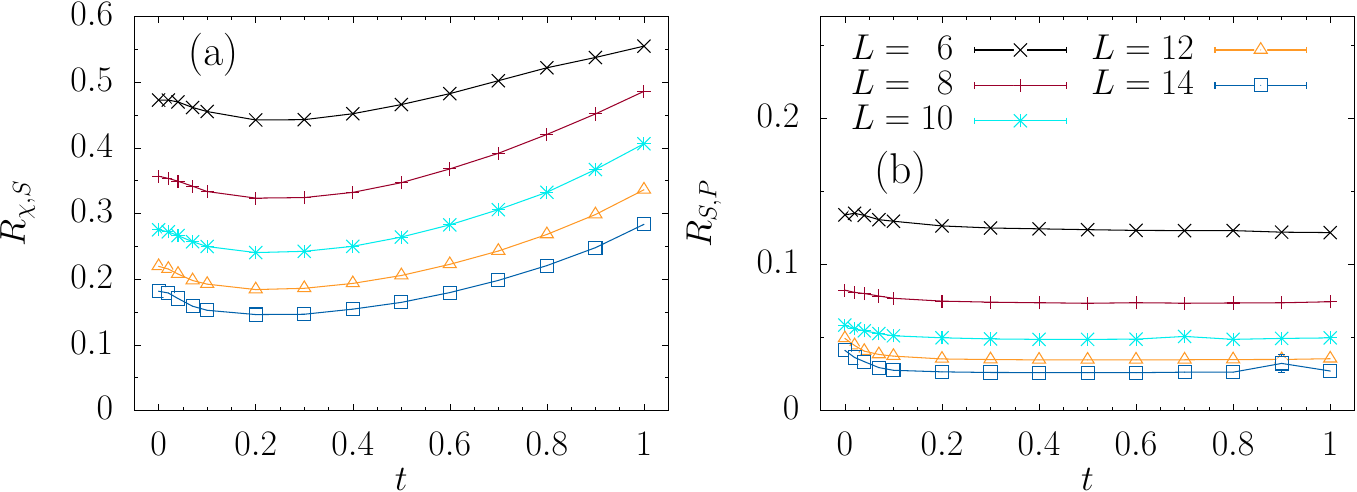}
  \caption{\label{fig:corr_w20}Correlation ratios for (a) spin susceptibility
    at $\bm{M}$ and (b) parity correlation function at $\bm{\Gamma}$. Here, $\omega_0=2.0$, $\beta=L$.}
\end{figure}

The derivative of the free energy with respect to $t$ is given by the average
kinetic energy
\begin{equation}\label{eq:dfdt}
  \frac{\partial F}{\partial t} = - \sum_b \langle\hat{K}_b\rangle \,.
\end{equation}
Results are shown in Fig.~\ref{fig:scal_w20}(b).  At $t=0$, where the model has
the local $\mathbb{Z}_2$ symmetry, the kinetic energy vanishes by symmetry and
only phonon-mediated hopping takes place.  We  can  expand the free  energy  around   $t=0$,
\begin{equation}
  F(t)  =   F_0  - t^2  \int_0^{\beta}    d \tau   \sum_b \langle \hat{K}_b(\tau)  \hat{K}_b(0) \rangle_0   +  {\cal O}(t^4)   
\end{equation}
with $F_0$ the free energy at $t=0$ and $\langle \hat{O}\rangle_0 $ the
expectation value of an observable $\hat{O} $ with respect to the Hamiltonian
with $t=0$. Here, only  even  powers  occur  since  $F(t) =  F(-t)$.   Hence,
$ {\partial F}/{\partial t} = - 2 t \int_0^{\beta} d \tau \sum_b \langle
\hat{K}_b(\tau) \hat{K}_b(0) \rangle_0 + {\cal O} (t^3)$.  Since the
time-displaced correlation function is positive, $ {\partial F}/{\partial t} $
decreases linearly with $t$ for small $t$.  This is consistent with
the QMC data. Within the numerical resolution and for our choice of $\beta=L$,
${\partial F}/{\partial t}$ is smooth as a function of $t$.    As mentioned  above, 
the  analytical  behavior  of  the  free energy  around $t=0$  implies that the
physics at small $t>0$ is adiabatically connected to that at $t=0$.

\begin{figure*}[ht]
\includegraphics[width=0.72\linewidth]{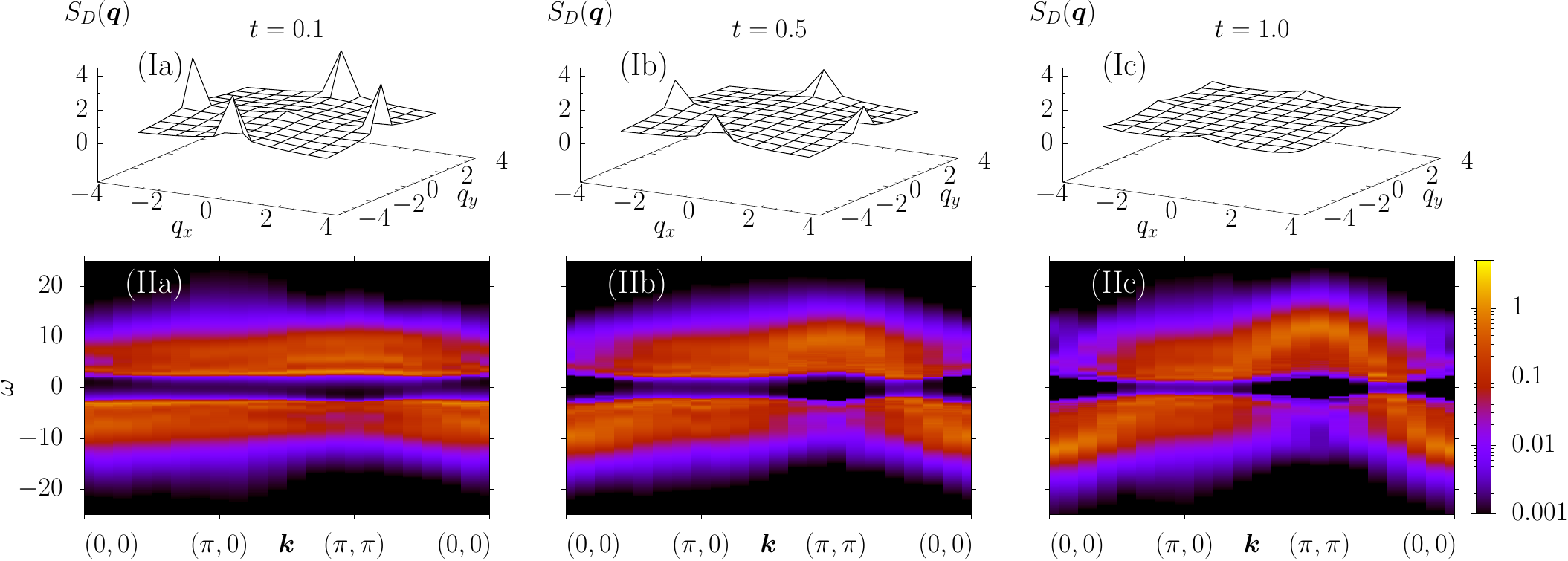} 
\caption{\label{fig:Spectral_Green_w20}(Ia)-(Ic) Equal-time dimer correlation
  function in the first Brillouin zone for $\beta=L=10$. (IIa)-(IIc)
  Single-particle spectral function $A(\bm{k},\omega)$ for $\beta=L=14$. Here, $\omega_0=2.0$.}
\end{figure*}

The results for the correlation ratios based on the spin susceptibility and the parity
equal-time correlation function in Fig.~\ref{fig:corr_w20} indicate the absence of spin order
for all values of $t$ considered. In Figs.~\ref{fig:Spectral_Green_w20}(Ia)-(Ic),
we present the dimer correlation function for
$t=0.1$, 0.5, and 1.0. For small $t$, where $\pi$-fluxes are dynamically
generated, the system is in the $(\pi,0)$ VBS phase. When the hopping is increased, the
dominant peaks at $(\pi,0)$ and equivalent wave vectors start to decrease and
the VBS order melts. At $t=1.0$, the correlation function is almost flat.

Although VBS order is suppressed with increasing $t$, the single-particle gap
remains open, as visible from the single-particle spectral function in
Figs.~\ref{fig:Spectral_Green_w20}(IIa)-(IIc). The latter evolves towards a
cosine band structure. The band width is determined by an
effective hopping
$t_{\mathrm{eff}}=t-({g}/{N_\text{b}})\sum_b \langle \hat{X}_b \rangle$ due to
the coupling to the phonons. At $t=1.0$ and $\omega_0=2.0$, we obtain
$t_{\mathrm{eff}}=2.25$. 

\begin{figure}[t]
\includegraphics[width=0.49\linewidth]{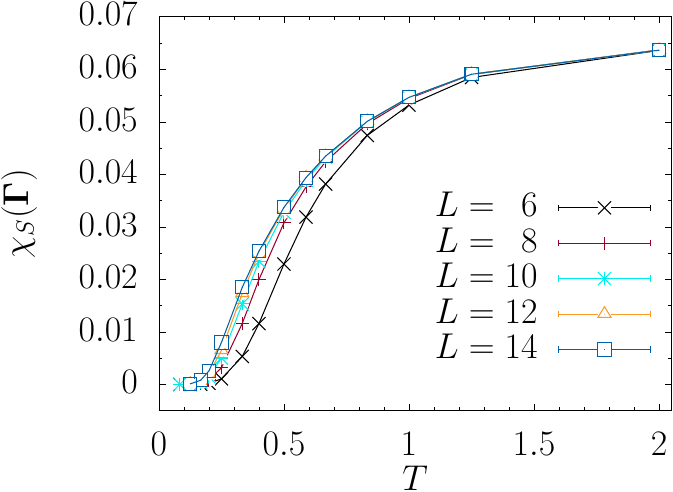}
  \caption{\label{fig:spin_suscep_w20}Spin susceptibility $\chi_S(\bm{\Gamma})$
    as a function of temperature. Here, $t=1.0$,  $\omega_0=2.0$.}
\end{figure}  

The pseudogap phase at $t=1$ corresponds to an O(4) symmetric finite-temperature
phase.  Furthermore, the uniform spin susceptibility in
Fig.~\ref{fig:spin_suscep_w20} supports the existence of a finite spin gap.  We
understand this phase in terms of preformed pairs that will order at lower
temperature.  At $t=1$, a $\pi$ flux is absent and the $f$-fermion
picture introduced above is no longer valid. Instead, we
interpret the results in terms of an instability of an underlying
$(\pi,\pi)$-nested Fermi surface.  In this case, and at the mean-field level
that becomes exact in the adiabatic limit, the transition temperature will
follow an essential singularity.  Hence, for our choice $\beta = L$, we expect
to be above the expected transition temperature, limiting our ability to draw
conclusions about the ground state. Strictly speaking, a $\beta=L$ scaling (\ie,
$z=1$) is no longer justified in the absence of dynamically generated $\pi$ fluxes.
Using lower temperatures within the present algorithm is challenging in this parameter
region due to long autocorrelation times. While it is unclear from the present
data which ordering wave vector is picked up for $T\to0$ at large $t$,
previous studies \cite{Xing21,Cai21,Goetz22,cai2022robustness,Feng22} suggest a
$(\pi,\pi)$-ordered VBS ground state.

\subsection{Flux crossover within the AFM phase}

\begin{figure}[b]
 \includegraphics[width=1\linewidth]{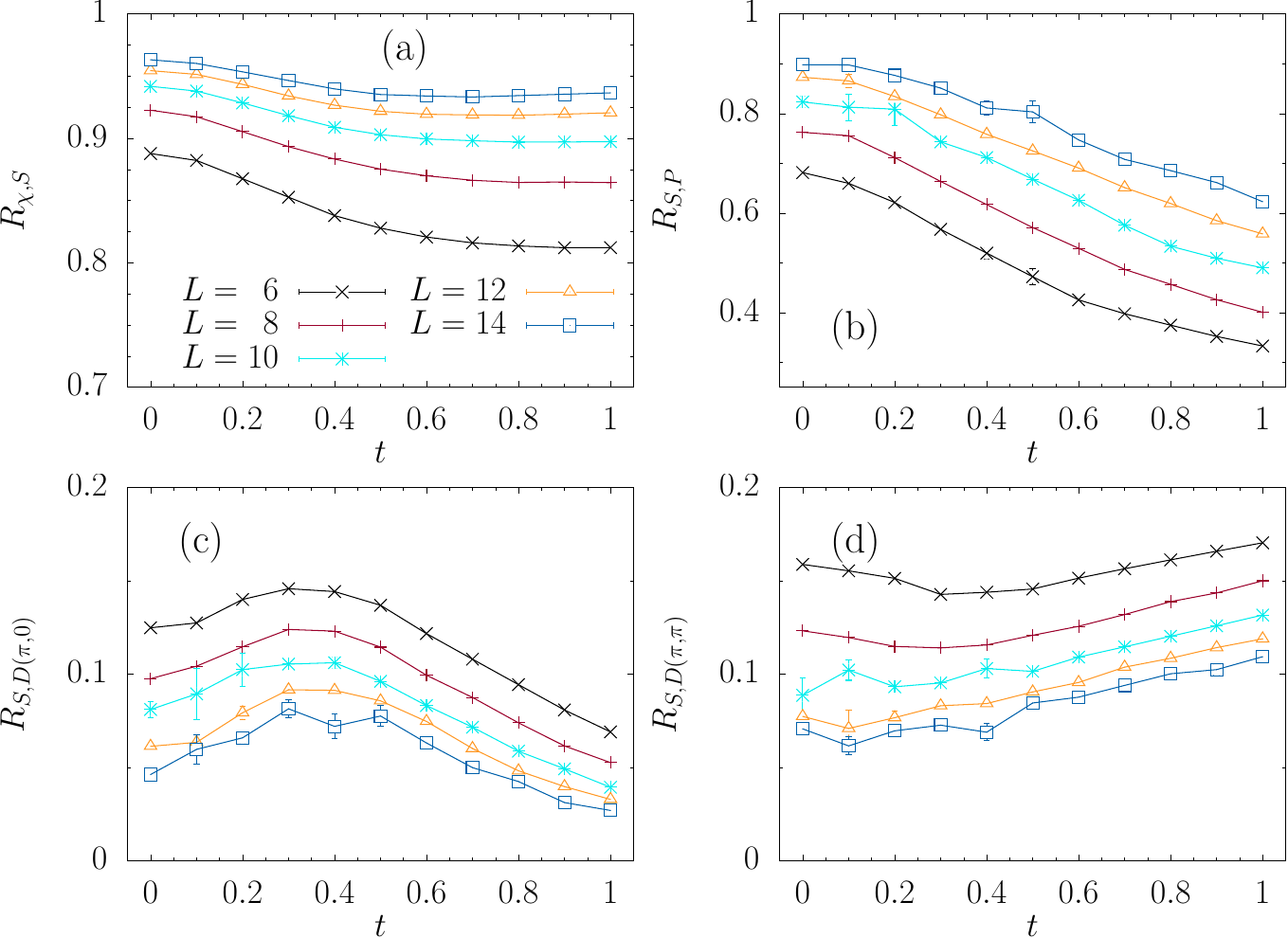}
 \caption{\label{fig:corr_w40}Correlation ratio of (a) the spin susceptibility
    at $\bm{M}$, (b) the parity correlations at $\bm{\Gamma}$, (c) dimer
    correlations at $\bm{X}$, (d) dimer correlations at $\bm{M}$. Here, 
    $\omega_0=4.0$, $\beta=L$. }
\end{figure}

In this section, we consider a value $\omega_0=4.0$, for which the system is in
the AFM phase according to Fig.~\ref{fig:phases}. AFM order is revealed by the
correlation ratios in Fig.~\ref{fig:corr_w40} for the entire range of hoppings
considered, $0\leq t\leq 1$. Specifically, long-range order is visible at
wave vector $\bm{M}$ in the spin sector [Fig.~\ref{fig:corr_w40}(a)] and at
$\bm{q}=\bm{\Gamma}$ in the parity sector [Fig.~\ref{fig:corr_w40}(b)].  In
contrast, the equal-time dimer correlation ratio excludes the presence of VBS
order at $\bm{M}$ and $\bm{X}$ [Figs.~\ref{fig:corr_w40}(c) and \ref{fig:corr_w40}(d)].  We note that
the AFM phase corresponds to a Lorentz invariant phase, so that irrespective of
a plaquette flux, the adopted $\beta = L $ scaling suffices to capture
ground-state properties.

\begin{figure}[t]
 \includegraphics[width=1\linewidth]{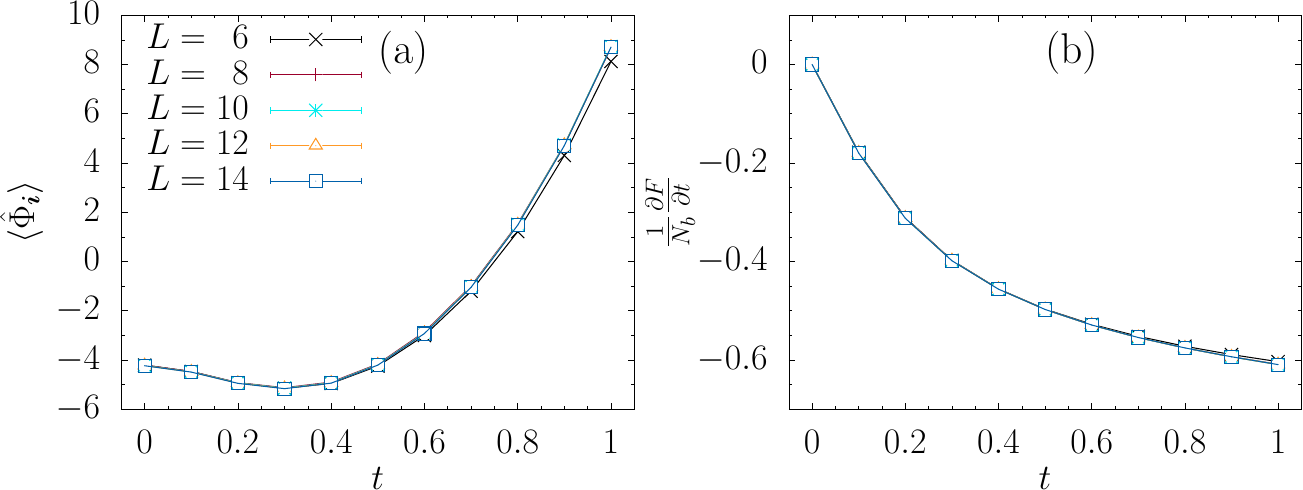}
  \caption{\label{fig:scal_w40}Flux and free-energy derivative with respect to $t$. Here,
    $\omega_0=4$, $\beta=L$. }
\end{figure}

Results for the flux as a function of $t$ [Fig.~\ref{fig:scal_w40}(a)] are
qualitatively very similar to those for $\omega_0=2.0$
[Fig.~\ref{fig:scal_w20}(a)]. The flux and the derivative of the
free energy with respect to $t$ are smooth at the considered resolution of $t$
[see Fig.~\ref{fig:scal_w40}(b)]. Comparison of Figs.~\ref{fig:scal_w20}(b) and \ref{fig:scal_w40}(b) reveals that the slope of ${\partial F}/{\partial
  t}$, $ \int_0^{\beta} d \tau \sum_b \langle \hat{K}_b(\tau) \hat{K}_b(0) \rangle_0 $,
is reduced for $\omega_0=4.0$. This is a consequence of the increased gap 
between different $\hat{Q}_{\i} $ sectors. 

A pure $\mathbb{Z}_2$ lattice gauge theory supports deconfined and confined
phases separated by an Ising transition at which fluxes (\ie, visons)
proliferate \cite{Wegner71,Fradkin13,Sachdev19}.  In the AFM phase, the fermions
are bound in particle-hole pairs that carry no $\mathbb{Z}_2$ charge.  Hence, the
AFM and gauge fluctuations effectively decouple at low energies, and two AFM
phases are possible. The gauge field is deconfined in the AFM* phase but
confined in the AFM phase. Such transitions have been discussed in Refs.~\cite{Gazit17,Gazit18}.
 
Because the hopping $t$ explicitly breaks the $\mathbb{Z}_2$ symmetry in our case,
the AFM* and AFM phases cannot be strictly distinguished in the sense that
they are separated by a critical point.  Nevertheless, we understand the sign
change in the flux in terms of a proliferation of visons and our data in
terms of an AFM* to AFM \textit{crossover}.

\section{Discussion and Conclusions}\label{sec:conclusion}

We studied a modified SSH model on a square lattice using an
auxiliary-field QMC approach inspired by Ref.~\cite{Seki18}.  By
adding a symmetry-allowed electronic
interaction, we were able to integrate out the phonon degrees of freedom in the
whole parameter space.  This results in imaginary-time correlations between
discrete auxiliary fields, which we sample with a combination of sequential
single-spin-flip and global updates.  In Appendix \ref{sec:appendix_comparison},
we argue that this discrete field approach is more efficient than updating the
continuous phonon fields.  With this method, we were able to study the
phase diagram of the model as a function of the phonon frequency and the hopping
strength.

In the original electron-phonon context of the SSH model, the ratio of
phonon-assisted hopping to direct hopping has to be small
\cite{Su79}.  However, allowing more general values of the hopping strength
provides a direct route between a model with dominant direct hopping and a model
with dominant phonon-assisted hopping. In the limiting case $t=0$, the symmetry
of the model is enhanced by a local $\mathbb{Z}_2$ symmetry and it maps on an
unconstrained lattice gauge theory. For any value of $t$, the model has a global O(4)
symmetry. As a result, an AFM phase is degenerate with CDW and SC ones,
and a partial particle-hole transformation maps  the   three AFM  order
parameters  onto one CDW and two SC order parameters.  These symmetries are also present
in the SSH model without the extra term. 
    
The limit $t=0$ is special due to the local $\mathbb{Z}_2$ symmetry.
However, since the free energy is not singular, the physics of $t=0$ is
representative of larger regions of the phase diagram.  At $t=0$, it is
convenient to adopt a slave-spin or orthogonal fermion representation [Eq.~(\ref{eq:fractionalize})], in which
the original $c$ fermion of the SSH model is fractionalized into an $f$ fermion
with electronic quantum numbers and an Ising spin $\tau$ that carries a $\mathbb{Z}_2$
charge. Whereas the $c$ fermion is localized, the $f$ fermions are itinerant and
subject to $\pi$ fluxes dynamically generated by the phonon degrees of
freedom. The fluxes cause the $f$ fermions to acquire a Dirac band
structure. Our simulations reveal an O(4) symmetric phase [a $(\pi,0)$-VBS
solid] that gives way to states with broken O(4) symmetry (\eg, the AFM phase)
at large $\omega_0$. The reduction of  O(4)  to
SO(4) corresponds to a finite-temperature
Ising transition in which the odd (AFM) or even (CDW/SC) parity sector is
spontaneously chosen.  At $T=0$, the SU(2) symmetry of the AFM or CDW/SC is
further reduced to U(1).

Our results suggest the transition from the $(\pi,0)$ VBS to the AFM/CDW/SC
phase, driven by the phonon frequency, is continuous.  The dynamical VBS
structure factor supports an emergent U(1) symmetry in the sense that it exhibits a
linear dispersion at criticality. Within the uncertainty, the VBS and spin velocities match at the critical 
point, as consistent
with emergent Lorentz invariance.  Finally, the Ising transition temperature
vanishes in  the  proximity of  the critical point. Overall, our data provide evidence for 
a DQCP, albeit in a model with $\mathrm{O}(4) \times \mathrm{C}_4$
symmetry, as opposed to $\mathrm{SU}(2) \times \mathrm{U}(1)$ \cite{Liu18} or
$\mathrm{SU}(2) \times \mathrm{C}_4$ \cite{Sandvik07}.  However, we also note
that other results point toward a weakly first-order
transition \cite{LiuZH23,HeYC23,Song23,Chen23}.  This does not impair our results on finite
lattices that exhibit signs of pseudo-criticality.  
Following  the  theory of  DQCP,    the  critical  point  is  described  by  a 
compact U(1)  gauge  theory   of  spinons.   The  authors of  Ref.~\cite{Seifert23} argue 
that such  a  state  exhibits  a Peierls  instability since  single-monopole  instances 
become  relevant.   It  is  hence intriguing to  repeat  our  calculations  with  modified 
model parameters so  as to lower  the  value  of  the  critical phonon frequency. If  single-monopole 
 instances  turn out  to be  relevant in the  adiabatic limit,  then   we   expect the  
transition to evolve  to  a  strong  first-order one.  

At small but finite values of $t$, the local $\mathbb{Z}_2$ symmetry holds
only on short time scales.  The phases that we observe in this regime are
remarkably similar to those in Ref.~\cite{Assaad16}, where the local
symmetry is exact.  This is an encouraging result for quantum simulations of
gauge theories, where it is often hard to impose the constraint on all time
scales.  Note, however, that in the considered parameter range, a spin-liquid
phase remains illusive.

The AFM phase with Lorentz invariant critical fluctuations (spin waves) observed
at large frequencies is robust to the vanishing of $\pi$ fluxes at large $t$.
On the other hand, and for the temperatures considered ($\beta=L$),
$(\pi,0)$ VBS order gives way to a pseudogap phase.  We understand the latter
in terms of thermal fluctuations of the $(\pi,\pi)$ VBS phase observed in this
parameter range at lower temperatures
\cite{Xing21,Cai21,Goetz22,cai2022robustness,Feng22}. It has a spin gap and, due
to the O(4) symmetry, identical charge and spin susceptibilities. A possible
interpretation is in terms of disordered singlets, whose dynamics is expected to
manifest itself as a non-vanishing specific heat.

The remarkable richness of the phase diagram motivates future
investigations.  Furthermore, the fact that we have an efficient discrete-field
representation of an SSH-type model provides the basis for several other directions.
For example, one can add a Hubbard term to break down the symmetry from O(4) to
SO(4).  Aside from differences in critical phenomena (\eg, the absence of a
finite-temperature Ising transition), we expect the phase diagram to remain
unchanged and hence robust to weak O(4) symmetry breaking.  Another interesting
direction is doping. In the phases with broken O(4) symmetry, we conjecture a
first-order spin-flop-like transition to a superconducting state upon
doping. The fate of the VBS state requires numerical investigation.

\begin{acknowledgments}
  The authors would like to thank K. Seki, N. C. Costa and J. Willsher for  interesting  discussions.
  The authors gratefully acknowledge the Gauss Centre for Supercomputing
  e.V. (www.gauss-centre.eu) for funding this project by providing computing
  time on the GCS Supercomputer SuperMUC-NG at the Leibniz Supercomputing Centre
  (www.lrz.de).  The authors gratefully acknowledge the scientific support and
  HPC resources provided by the Erlangen National High Performance Computing
  Center (NHR@FAU) of the Friedrich-Alexander-Universit\"at Erlangen-N\"urnberg
  (FAU) under NHR project 80069. NHR funding is provided by federal and
  Bavarian state authorities. NHR@FAU hardware is partially funded by the German
  Research Foundation (DFG) through grant 440719683.  F.F.A. thanks the W\"urzburg-Dresden
  Cluster of Excellence on Complexity and Topology in Quantum Matter ct.qmat
  (EXC 2147, project-id 390858490) as well as the DFG under the grant AS 120/16-1 (Project number 493886309) that is part of the collaborative research project SFB Q-M\&S funded by the Austrian Science Fund (FWF) F 86. F.F.A. and A.G. thank the DFG for financial support under the grant AS 120/19-1  (Project number 530989922).
\end{acknowledgments}


\appendix 

\section{Trotter decomposition}\label{sec:appendix_trotter}

In this appendix, we compare the Trotter decomposition used to obtain the results
in the main text with an asymmetric decomposition scheme.  By splitting the
exponential of the Hamiltonian asymmetrically, 
\begin{equation}\label{eq:trot_asymm}
  e^{-\Delta \tau \hat{H} } = 
  e^{-\Delta \tau \hat{H}_{\omega_0} } \prod_b e^{-\Delta\tau \hat{H}_{\lambda,b}} 
  e^{- \Delta \tau  \hat{H}_t }   + \mathcal{O}(\Delta\tau^{2}) \,,
\end{equation}
the Hermiticity of the time propagation is lost.
The formulation of the partition function in Eq.~(\ref{eq:Z}) remains valid if
we drop the summation over the index $\alpha$ and instead define
\begin{equation}
  v_{b,\tau}= g \sqrt{\frac{\Delta \tau}{4 \lambda}}  \eta(l_{b,\tau-1}) 
\end{equation}
and
\begin{equation}
  B(\tau_1,\tau_2) = \prod_{\tau=\tau_2+\Delta \tau}^{\tau_1}   \left(\prod_{b} e^{\sqrt{\Delta\tau\lambda}\eta(l_{b,\tau})K_b} \right) e^{\Delta \tau  t \sum_b  K_b} \,.
\end{equation}

\begin{figure}[b]
 \includegraphics[width=\linewidth]{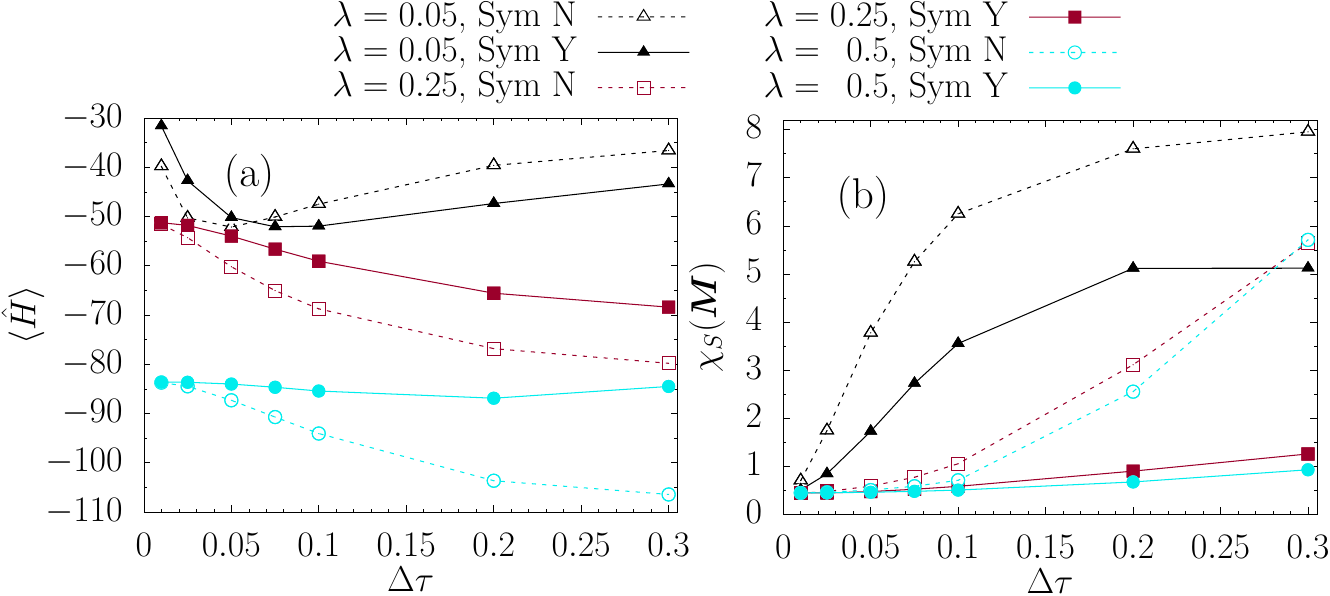}
  \caption{\label{fig:Trot}Energy $\langle \hat{H} \rangle$ and spin
    susceptibility $\chi_S(\bm{M})$ as a function of the Trotter step size at
    different $\lambda$ and for an asymmetric (Sym N) or symmetric (Sym Y) Trotter
    decomposition. Here, $t=0.1$, $\omega_0=2.0$, $\beta=L=6$.}
\end{figure}

When measuring observables, the results of the decomposition schemes of
Eqs.~(\ref{eq:symmZ}) and (\ref{eq:trot_asymm}) scale with $\Delta
\tau^2$. Naively, one would expect a linear scaling for the asymmetric
decomposition, but the term linear in $\Delta \tau$ vanishes under the
assumption that all operators in the decomposition and the observable are real
representable, as shown by Fye~\cite{Fye86}. However, Fig.~\ref{fig:Trot} shows
that the prefactor for the symmetric decomposition is much smaller than that
for the asymmetric decomposition.

In Fig.~\ref{fig:Trot}, we present the average energy $\langle \hat{H} \rangle$
and the spin susceptibility $\chi_S(\bm{M})$ as a function of the Trotter step
size $\Delta \tau$ for both decomposition schemes and for different
values of the electronic coupling strength $\lambda$. Surprisingly, for small
$\lambda$, the energy does not converge to a finite value as $\Delta \tau$ is
decreased [Fig.~\ref{fig:Trot}(a)]. We expect that in the limit
$\Delta \tau\rightarrow 0$ the systematic error due to the Trotter decomposition
scales to zero and the results of both decomposition schemes extrapolate to the
same value. However, for $\lambda=0.05$, it is not clear to which value the energy
extrapolates. By increasing $\lambda$, the results converge to approximately the
same finite value. However, the energy calculated with the symmetric version
converges faster. Similar behavior is observed for the spin susceptibility in
Fig.~\ref{fig:Trot}(b).

\section{Comparison with other QMC approaches}\label{sec:appendix_comparison}

In this appendix, we provide a short comparison of our method with other
approaches. First, we compared it to an algorithm that does not make use of
integrating out the phonons and is based on a discrete Hubbard-Stratonovitch
decomposition to decouple the $\lambda \hat{K}_b^2$ term,
\begin{equation}
  e^{\Delta \tau \lambda \hat{K}_b^2}  = \frac{1}{4} \sum_l \gamma(l) e^{\sqrt{\Delta \tau \lambda} \eta(l) \hat{K}_b} \, .
\end{equation}
The partition function can be written as
\begin{eqnarray}\label{eq:not_integrated}
  Z &=& \sum_{\left\{l_{b,\tau}\right\}} \left(\prod_{b,\tau} \frac{ \gamma(l_{b,\tau})}{4} \right) 
  \int \prod_{b,\tau} d\x_{b,\tau}  \det\left[ 1+ B(\beta,0) \right]  \nonumber \\
&& \quad \times   e^{ - \Delta \tau \sum_{b,\tau} \left[ \frac{m}{2} \left(\frac{\x_{b,\tau+1}-\x_{b,\tau}}{\Delta \tau} \right)^2 + \frac{k}{2} \x_{b,\tau}^2 \right] }
\end{eqnarray}
with the matrix
\begin{eqnarray}
  B(\tau_1,\tau_2) &=& \prod_{\tau=\tau_2+\Delta \tau}^{\tau_1}    \left(\prod_{b} e^{- \Delta \tau g \x_{b,\tau} K_b} \right) \nonumber \\
&&  \times   \left(\prod_{b} e^{\sqrt{\Delta\tau\lambda}\eta(l_{b,\tau})K_b} \right) e^{\Delta \tau  t \sum_b  K_b} \, .
\end{eqnarray}
In this case, the stochastic sampling is over the discrete Hubbard-Stratonovitch fields
$\{\eta(l_{b,\tau})\}$ and the phonon fields $\{\x_{b,\tau}\}$. We used
single-spin-flip updates with a Metropolis-Hastings acceptance-rejection step
for both kinds of fields.

Motivated by Ref.~\cite{Goetz22}, we also used a Langevin-based algorithm for
comparison. In order to employ the Langevin updating scheme, we
decoupled the electron interaction with a continuous Hubbard-Stratonovitch transformation,
\begin{equation}
  e^{\Delta \tau \lambda \hat{K}_b^2}  = \frac{1}{\sqrt{2 \pi}} \int d\phi  e^{-\frac{1}{2} \phi^2 - \sqrt{2 \Delta \tau \lambda} \phi \hat{K}_b} \, .
\end{equation}
In this case, the partition function is given by
\begin{eqnarray}\label{eq:Z_Langevin}
  Z &\propto& \int \left( \prod_{b,\tau} d\phi_{b,\tau} d\x_{b,\tau} \right)   \times  \det\left[ 1+ B(\beta,0) \right] \\
&& \quad  \times e^{ - \Delta \tau \sum_{b,\tau} \left[ \frac{m}{2} \left(\frac{\x_{b,\tau+1}-\x_{b,\tau}}{\Delta \tau} \right)^2 + \frac{k}{2} \x_{b,\tau}^2 \right] } \nonumber
\end{eqnarray}
with
\begin{eqnarray}
  B(\tau_1,\tau_2) &=& \prod_{\tau=\tau_2+\Delta \tau}^{\tau_1}    \left(\prod_{b} e^{- \Delta \tau g \x_{b,\tau} K_b} \right)  \nonumber \\ 
& & \times    \left(\prod_{b} e^{-\sqrt{2\Delta\tau\lambda}\phi_{b,\tau} K_b} \right) e^{\Delta \tau  t \sum_b  K_b} \, .
\end{eqnarray}
For an introduction on the Langevin updating scheme, see
Refs.~\cite{Batrouni85,Gardiner09,Batrouni19} and references therein.

\begin{figure}[t]
 \includegraphics[width=\linewidth]{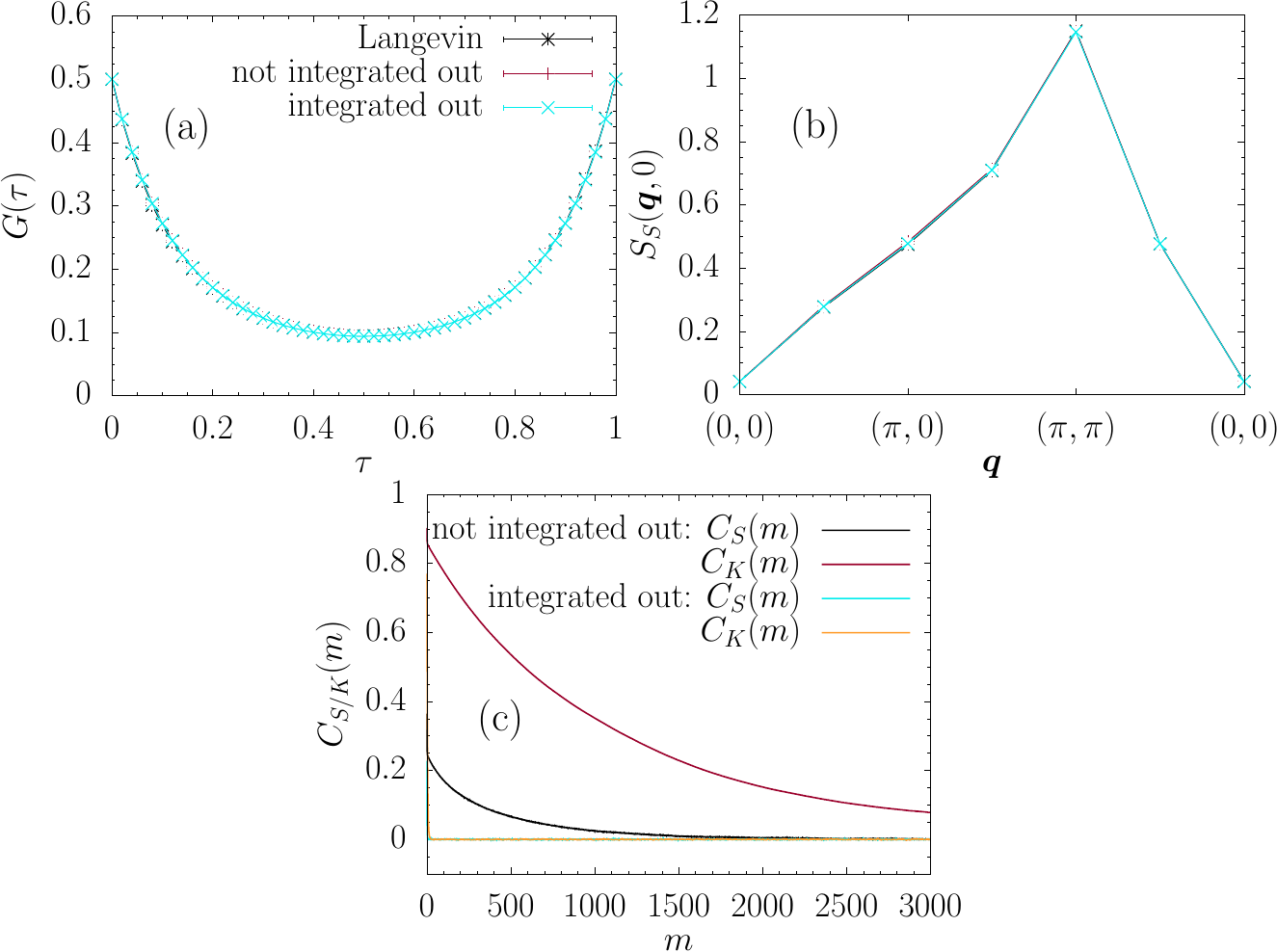}
  \caption{\label{fig:comp}(a) Local imaginary-time Greens function $G(\tau)$ and
    (b) spin correlation function $S_S(\bm{q},0)$ in the first Brillouin
    zone for $L=4$, $\beta=1.0$, $t=1.0$, $\omega_0=3.0$, and $\Delta \tau=0.02$. (c)
    Correlation functions $C_S(m)$ and $C_K(m)$ as a function of the number of
    sweeps $m$ for the same parameters but with simulations 
    performed on a single core.  ``Integrated out'' means we used the algorithm
    based on the partition function of Eq.~(\ref{eq:Z}), ``not integrated out''
    refers to simulations based on Eq.~(\ref{eq:not_integrated}), and
    ``Langevin'' to the use of Eq.~(\ref{eq:Z_Langevin}).}
\end{figure}

In Figs.~\ref{fig:comp}(a) and (b), we compare the local imaginary-time Greens
function $G(\tau)$ and the spin correlation function $S_S( \bm{q},0)$ for all
three methods. The results are in good agreement.  In general, we noticed a
reduction of the autocorrelation time of several observables if the phonons are
integrated out. In Fig.~\ref{fig:comp}(c), we compare the autocorrelation time
of the equal-time spin correlation function $S_S(\bm{M},0) $ and the average
kinetic energy $\langle\hat{H}_t\rangle=-t\sum_b \langle \hat{K}_b \rangle$ for
the methods based on Eqs.~(\ref{eq:Z}) and (\ref{eq:not_integrated}),
respectively. The correlation function is defined by
\begin{equation}
  C_{\hat{O}}(m)=  \frac{\sum_{i=1}^{N_{\mathrm{Bin}} - m}(O_i -  \langle \hat{O} \rangle) (O_{i+m} - \langle \hat{O} \rangle)}{\sum_{i=1}^{N_{\mathrm{Bin}} - m}(O_i -  \langle \hat{O} \rangle)^2}
\end{equation}
with
\begin{equation}
  \langle\hat{O} \rangle = \frac{1}{N_{\mathrm{Bin}} } \sum_{i=1}^{N_{\mathrm{Bin}} } O_i \,,
\end{equation}
where $O_i$ is the value of the observable in the $i$th bin and
$N_{\mathrm{Bin}}$ is the total number of measurements for a single run. The
shorter the autocorrelation time, the faster the correlation function drops to
zero, indicating uncorrelated measurements after a certain number of sweeps. A
sweep is defined here as visiting every auxiliary field twice and
proposing an update with a Metropolis-Hastings acceptance-rejection step. For
Fig.~\ref{fig:comp}, we collected around $6\times10^6$ sweeps on a single core. The
correlation functions for the method with the phonons integrated out need on the
order of ten sweeps to drop to zero, whereas for the other method on the order
of $10^{3}$ sweeps are required. Comparing the autocorrelation time with the
Langevin method is difficult because the notion of an update is different.

The method used in the main text can be successfully used at higher
phonon frequencies compared to the Langevin method because of the negative
impact of zeros in the determinant on the latter [see also Ref.~\cite{Goetz22}].

\section{Measuring phonon observables}\label{sec:appendix_phonon}

Because we integrated out the phonons in the action, we cannot directly access
observables that are functions of phonon variables. To circumvent this
issue, we introduced a source term in the phonon action,
\begin{equation}
  S_j = - \int_0^{\beta} d \tau \sum_b j_b(\tau) \x_b(\tau) = -\sum_{b,\tau} \Delta \tau j_{b,\tau} \x_{b,\tau} \,.
\end{equation}
This allows us to formulate the expectation value of a phonon displacement
operator as the derivative of the action with respect to the variable $j$
\cite{Hewson01,Weber15}:
\begin{equation}
  \langle \hat{X}_{b_1,\tau_1} \rangle = \frac{1}{\Delta\tau} \frac{\partial \ln Z}{\partial j_{b_1,\tau_1}}  \Big{|}_{\{j_{b,\tau}=0\}}
  = -\frac{1}{2} \sum_\tau (A^{-1})_{\tau_1,\tau} \langle v_{b_1,\tau}\rangle \,.
\end{equation}
A similar relation holds for the expectation value of the product of two phonon fields,
\begin{eqnarray}
&&  \langle \hat{X}_{b_1,\tau_1} \hat{X}_{b_2,\tau_2} \rangle = \frac{1}{\Delta\tau^2} \frac{1}{Z} \frac{\partial^2  Z}{\partial j_{b_1,\tau_1}\partial j_{b_2,\tau_2}}  \Big{|}_{\{j_{b,\tau}=0\}}  \\
&&  =  \frac{1}{2}\delta_{b_1,b_2}(A^{-1})_{\tau_1,\tau_2}   +\frac{1}{4} \sum_{\tau,\tau'} (A^{-1})_{\tau_1,\tau} (A^{-1})_{\tau_2,\tau'} \langle v_{b_1,\tau}  v_{b_2,\tau'}\rangle \,. \nonumber
\end{eqnarray}

\section{Additional data}\label{sec:appendix_data}

We consider a fixed value $t=1.0$ and vary the phonon frequency
$\omega_0$. According to Fig.~\ref{fig:phases}, we expect AFM order at large
$\omega_0$ and a disordered state (the pseudogap phase) at small $\omega_0$.

\begin{figure}[t]
 \includegraphics[width=\linewidth]{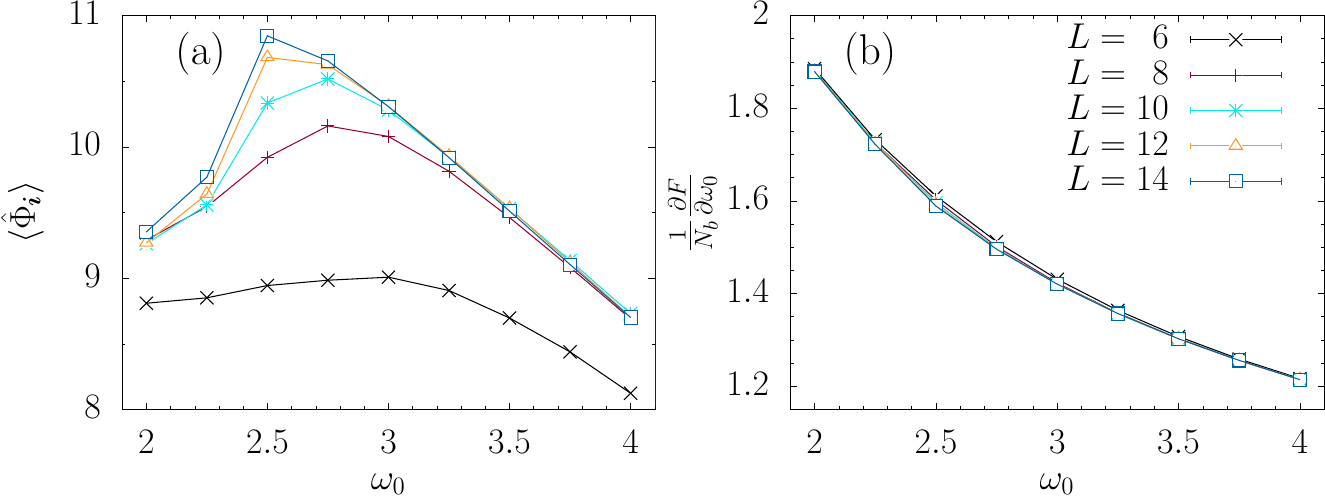}
  \caption{\label{fig:scal_t10}(a) Flux and (b) derivative of the free energy
    with respect to $\omega_0$. Here,  $t=1.0$, $\beta=L$. See also Fig.~\ref{fig:df_dw}(b) in erratum.}
\end{figure}

The average flux remains positive for the parameters considered, with a maximum
near $\omega_0=2.5$ [Fig.~\ref{fig:scal_t10}(a)]. The derivative of the
free energy with respect to the phonon frequency decreases smoothly with
increasing $\omega_0$ [Fig.~\ref{fig:scal_t10}(b)].

\begin{figure}[b]
 \includegraphics[width=\linewidth]{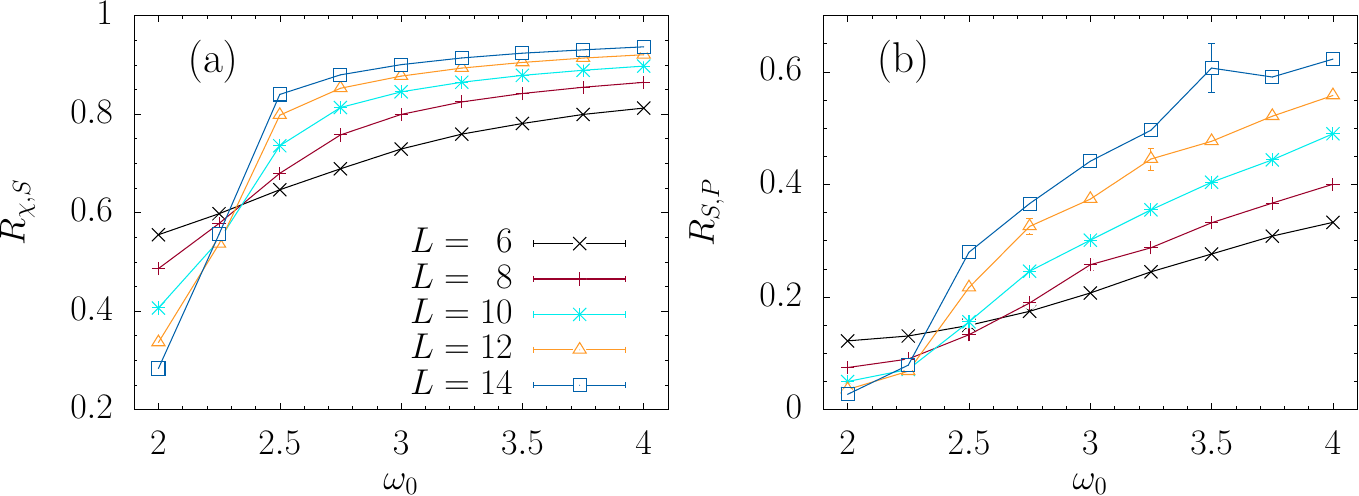}
  \caption{\label{fig:corr_t10}Correlation ratio for (a) spin susceptibility
    at $\bm{M}$ and (b) parity correlation function at $\bm{\Gamma}$. Here, $t=1.0$, $\beta=L$. }
\end{figure}

The correlation ratios of the spin susceptibility and the parity equal-time correlation function at the relevant
wave vector ($\bm{M}$ and $\bm{\Gamma}$, respectively) indicate ordering at
high phonon frequencies and a phase transition at a critical phonon frequency [Fig.~\ref{fig:corr_t10}].

\begin{figure*}[ht]
\includegraphics[width=0.72\linewidth]{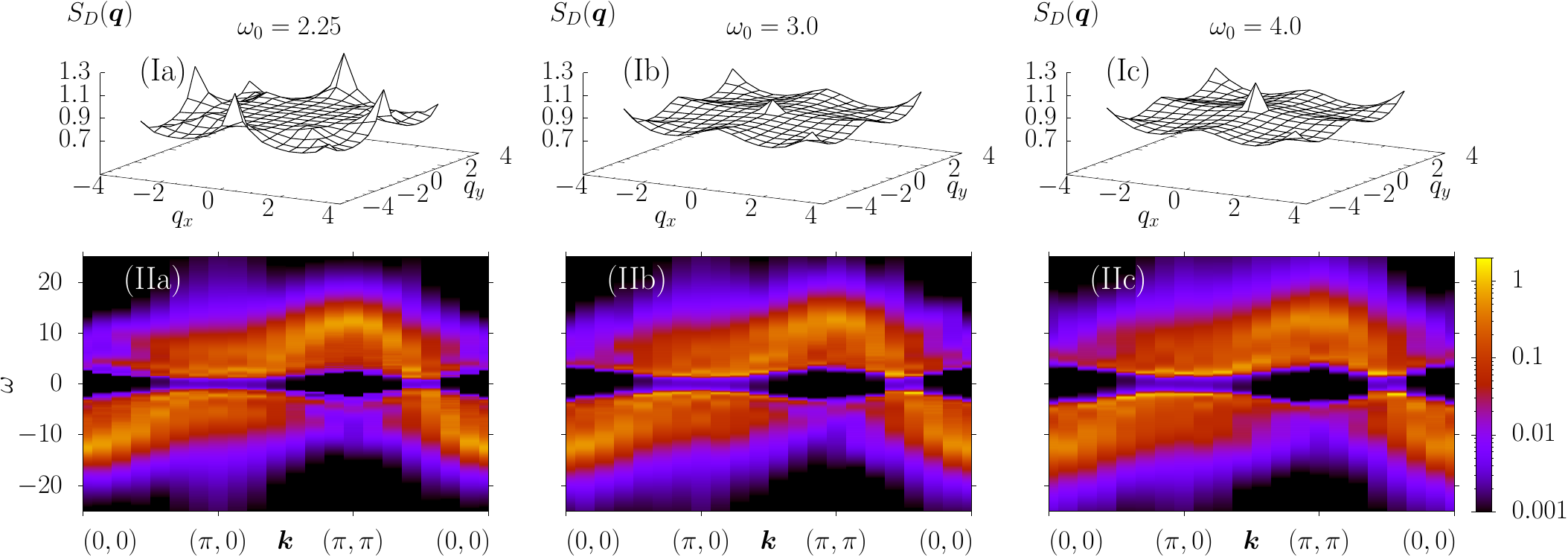} 
  \caption{\label{fig:Spectral_Green_t10}(Ia)-(Ic) Equal-time dimer correlation
    function in the first Brillouin zone, (IIa)-(IIc) single-particle
    spectral function $A(\bm{k},\omega)$. Here, $t=1.0$, $\beta=L=14$.}
\end{figure*}

Figure~\ref{fig:Spectral_Green_t10} shows the equal-time dimer correlation
function within the first Brillouin zone at different values of $\omega_0$ as
well as the single-particle spectral function $A(\bm{k},\omega)$.
The dimer
correlation function shows no ordering wave vector in the AFM phase and  no
order develops upon lowering the phonon frequency. However, 
with  decreasing  phonon frequency,  $(\pi,0)   $   VBS   fluctuations  grow,     thereby  
signaling   enhanced  proximity  to the    $(\pi,0)   $  ordered VBS  phase.   This observation is in agreement with
the results of Sec.~\ref{sec:low_w}. At the same time, the gap in the
single-particle spectral function remains open upon crossing the critical phonon
frequency. Spectral weight accumulates around the edges of the gap with increasing $\omega_0$.

\bibliography{paper}

\newpage

\onecolumngrid
\section*{Erratum}
Equation~(\ref{eq:df_dw}) of the original paper should be replaced by 
\begin{equation}
      \frac{\partial F}{\partial \omega_0}  = m \omega_0 \sum_b \langle \hat{X}_b^2 \rangle + \frac{g}{\omega_0} \sum_b \langle \hat{X}_b \hat{K}_b \rangle \,.
\end{equation}
This equation can be demonstrated by a canonical transformation of the position $\hat{X}_b \rightarrow \alpha \hat{\tilde{X}}_b$ and momentum $\hat{P}_b \rightarrow \frac{1}{\alpha} \hat{\tilde{P}}_b$ operators of the phonons. The canonical commutation relations $[\hat{X}_b,\hat{P}_{b'}] = [\hat{\tilde{X}}_b, \hat{\tilde{P}}_{b'}]$ and the partition function $
Z=\Tr e^{-\beta \hat{H}} = \Tr e^{-\beta \hat{H}(\alpha)}=Z(\alpha)
$ are invariant under this transformation, where we have used the following definition
\begin{equation}
\hat{H}(\alpha) =  \sum_{b}  (-t +  g\alpha\hat{\tilde{X}}_{b})  \hat{K}_b  - \lambda \sum_b \hat{K}_b^{2}  +  
\sum_b \left(   \frac{1}{2m\alpha^2} \hat{\tilde{P}}_b^2  +   \frac{k \alpha^2}{2}  \hat{\tilde{X}}_{b}^2 \right) \,.
\end{equation} 
As  apparent,  the electron-phonon coupling, $g^2/2k$,  as well as the phonon frequency,  $\omega_0  = \sqrt{k/m}$,  are   independent of  $\alpha$ and constitute the  two independent parameters  required to fully  define the model. Here we will vary the phonon frequency while keeping the  electron-phonon coupling  constant. 
Varying the partition function with respect to the parameter $\alpha$ yields the following relation
\begin{equation}
\frac{\partial \ln{Z(\alpha)}}{\partial \alpha}\big{|}_{\alpha=1} = \sum_b \langle g \hat{X}_b \hat{K}_b - \frac{1}{m} \hat{P}_b^2 + k \hat{X}_b^2\rangle = 0 = \frac{\partial \ln{Z}}{\partial \alpha}\,,
\end{equation}
which can be used in the calculation of the derivative of the free energy with respect to the phonon frequency $\omega_0$
\begin{equation}
\frac{\partial F}{\partial\omega_0} = - \frac{2m}{\omega_0} \frac{\partial F}{\partial m}  =   \frac{1}{\omega_0 m} \sum_b \langle  \hat{P}_b ^2\rangle = 
\frac{1}{\omega_0} \sum_b \langle g \hat{X}_b \hat{K}_b + k \hat{X}_b^2\rangle \,.
\end{equation}

\begin{figure}[h]
 \includegraphics[width=1\linewidth]{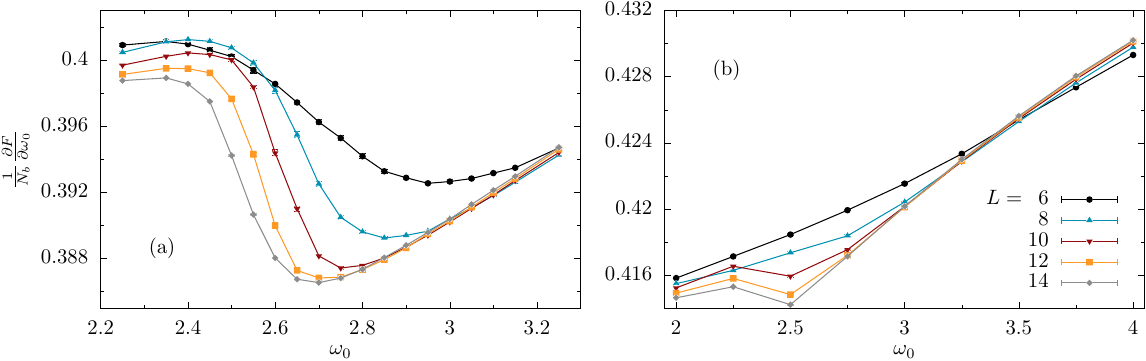}
  \caption{\label{fig:df_dw}Normalized free-energy derivative with respect to $\omega_0$ at (a) $t=0.1$ and (b) $t=1.0$. Here, $\beta = L$. }
\end{figure}

As a consequence, two figures of the original paper have to be replaced. Figure~\ref{fig:df_dw}(a) replaces Fig.~\ref{fig:scal_t01}(b) of the original paper and figure~\ref{fig:df_dw}(b) replaces Fig.~\ref{fig:scal_t10}(b) of the original paper.
Around the critical phonon frequencies, the derivatives change their qualitative behaviour, indicating the phase transitions. Importantly, the derivative at $t=0.1$ remains smooth on the considered scale and on the finite lattices [Fig.~\ref{fig:df_dw}(a)], where we argued for a deconfined quantum critical point or a weakly first order transition between a $(\pi,0)$-ordered valence bond solid and an antiferromagnetic phase in the original paper.  In Fig.~\ref{fig:df_dw}(b),  the  dip in  $\frac{1}{N_b} \frac{\partial F}{\partial \omega_0} $ reflects  the crossover from the  antiferromagnetic to pseudogap phase.

\end{document}